%% file: main.tex
\documentclass[manuscript,screen]{acmart}
\usepackage{paralist}
\usepackage{booktabs}
\usepackage[linesnumbered,noend, noline]{styles/algorithm2e}
\usepackage{tikz}
\usepackage{styles/pgf-pie}
\usetikzlibrary{positioning,shadows}
\newif\ifpienumberinlegend
\pgfkeys{/number in legend/.code=
    \expandafter\let\expandafter\ifpienumberinlegend
    \csname if#1\endcsname
    \ifpienumberinlegend

    \def\beforenumber##1\afternumber{}%
    \fi,
    /number in legend/.default=true
}

\definecolor{1c1}{RGB}{188,162,6}
\definecolor{1c2}{RGB}{137,129,80}
\definecolor{1c3}{RGB}{239,167,31}
\definecolor{1c4}{RGB}{88,194,241}
\definecolor{1c5}{RGB}{6,180,188}

\tikzset{mynode/.style={draw=white,solid,circle,fill=green,inner sep=1pt, thick,
text=black}}
\tikzset{arrow line/.style={dashed, line width= 2.5pt, color=#1}}
\usepackage{subfig}

\usepackage{algpseudocode}
\usepackage{url}

\makeatletter
\g@addto@macro{\UrlBreaks}{\UrlOrds}
\makeatother
\usepackage{breakurl}
\newcommand{\urls}[1]{{\scriptsize\url{#1}}}

\def\it{\textit}
\def\tr{\textrm}

\def\tt{\ft}

\usepackage{wrapfig}

\def\bf{\textbf}
\def\eq {Equation~}

\def\fig {Figure~}

\newcommand{\nd}{\vspace{1mm}\noindent}

\usepackage{balance}  
\usepackage{graphics} 
\usepackage{booktabs}
\usepackage{ccicons}

\usepackage{url}
\usepackage{fancybox}
\usepackage{multirow}
\usepackage{flushend}
\usepackage{booktabs}
\usepackage{tabularx}
\usepackage{comment}
\usepackage{array}
\usepackage[flushleft]{threeparttable}
\usepackage{mdframed}
\graphicspath{{}{images/}{dia/}}
\DeclareGraphicsExtensions{.pdf,.png}

\usepackage{listings}
\usepackage{hyperref}

\newcounter{o}
\newcounter{d}
\newcounter{t}
\usepackage{tikz}
\definecolor{1c1}{RGB}{188,162,6}
\definecolor{1c2}{RGB}{137,129,80}
\definecolor{1c3}{RGB}{239,167,31}
\definecolor{1c4}{RGB}{88,194,241}
\definecolor{1c5}{RGB}{6,180,188}

\tikzset{mynode/.style={draw=white,solid,circle,fill=green,inner sep=1pt, thick,
text=black}}
\tikzset{arrow line/.style={dashed, line width= 2.5pt, color=#1}}

\def\bf{\textbf}
\def\eq {Equation~}

\def\fig {Figure~}

\def\tbl {Table~}

\def\sec {Section~}
\def\secs {Sections~}

\def\it{\textit}
\def\tr{\textrm}
\def\tt{\mct}
\newcommand{\mct}[1]{{\footnotesize {\texttt {#1}}}}

\usepackage{paralist}
\usepackage{hyperref}
\usepackage{soul}

\normalsize

\usepackage{enumitem}

\usepackage{caption}
\usepackage{tikz}
\newcommand*\circled[1]{\tikz[baseline=(char.base)]{
            \node[shape=circle,draw,inner sep=1pt] (char) {#1};}}

\usepackage{rotating}
\usepackage{pdflscape}
\lstdefinestyle{inlinecode}{basicstyle={\ttfamily\scriptsize\bfseries}}

\usepackage{tcolorbox}
\newcommand{\emt}[1]{\emph{``#1''}}

\usepackage{paralist}
\usepackage[outercaption]{sidecap}    

\newcommand{\rev}[1]{\textcolor{black}{ #1}}
\usepackage{enumitem}

\newtcolorbox
{mybox}[2][]{colbacktitle=red!10!white,
colback=blue!10!white,coltitle=black!70!black,
title={#2},fonttitle=\bfseries,#1}

\AtBeginDocument{%
  \providecommand\BibTeX{{%
    \normalfont B\kern-0.5em{\scshape i\kern-0.25em b}\kern-0.8em\TeX}}}

\begin{document}

\title{Automatic API Usage Scenario Documentation from Technical Q\&A Sites}

\author{Gias Uddin}
\email{gias.uddin@ucalgary.ca}
\affiliation{
  \institution{University of Calgary}
  \country{Canada}
}
\author{Foutse Khomh}
\email{foutse.khomh@polymtl.ca}
\affiliation{
  \institution{Polytechnique Montr\'{e}al}
  \country{Canada}
}
\author{Chanchal K Roy}
\affiliation{%
  \institution{University of Saskatchewan}
  \country{Canada}
}
\email{croy@cs.usask.ca}

\renewcommand{\shortauthors}{Uddin et al.}

\begin{abstract}
The online technical Q\&A site Stack Overflow (SO) is popular among developers 
to support their coding and diverse development needs. 
To address shortcomings in API official 
documentation resources, several research has thus focused on augmenting official API documentation with insights (e.g., code examples) from 
SO. The techniques propose to add code examples/insights about APIs into its official documentation. 
Recently, surveys of software developers find that developers in SO consider 
the combination of code examples and reviews about APIs as a form of API documentation, and that they consider such a combination 
to be more useful than official API documentation when the official resources can be incomplete, ambiguous, incorrect, and outdated.  Reviews are 
opinionated sentences with positive/negative sentiments. However, 
we are aware of no previous research that attempts to automatically produce API documentation from 
SO by considering both API code examples and reviews. 
In this paper, we present two novel algorithms that can be used to automatically produce API 
documentation from SO by combining code examples and reviews towards those 
examples. The first algorithm is called statistical documentation, which shows the distribution of positivity and negativity around the code examples of an API using 
different metrics (e.g., star ratings). The second algorithm is called concept-based documentation, which clusters similar and conceptually relevant usage scenarios. 
An API usage scenario contains a code example, 
a textual description of the underlying task addressed by the code example, and the reviews (i.e., opinions with positive and negative sentiments)
 from other developers towards the code example. We deployed the algorithms in Opiner, a web-based platform to aggregate information about APIs from online forums.
We evaluated the algorithms by mining all Java JSON-based posts in SO and by conducting three user studies based on produced documentation from the posts. 
The first study is a survey, where we asked the participants to compare our proposed algorithms against a Javadoc-syle  documentation format (called as Type-based documentation in Opiner). 
The participants were asked to compare along four development scenarios (e.g., selection, documentation). The participants preferred 
our proposed two algorithms over type-based documentation. 
In our second user study, we asked the participants to complete four coding tasks using Opiner and the API official 
and informal documentation resources. The participants were more effective and accurate while using Opiner. In a subsequent survey, 
more than 80\% of participants asked the Opiner documentation platform to be 
 integrated into the formal API documentation to complement and improve the API official documentation.
\end{abstract}


\maketitle

\input{intro.tex}
\input{researchContext.tex}
\input{summarizationAlgos.tex}
\input{evalAgainstTypeSummary.tex}
\input{evalCoding}

\input{discussions}

\input{threats}
\input{related-work}
\input{summary}
\bibliographystyle{abbrv}
\bibliography{consolidated}
\end{document}

%% file: intro.tex
\section{Introduction}\label{sec:introduction}
APIs (Application Programming Interfaces) offer interfaces to reusable software components~\cite{Robillard-APIProperty-IEEETSE2012}.
Modern day rapid software development is facilitated by the numerous open source APIs that are available for any given task. 
Such is the popularity of APIs that the number of open source repositories in GitHub now is 100 million, an exponential 
increase over 67 million from only two years ago~\cite{website:github-octoverse}. With the growing number of open source repositories 
and the APIs supported by the repositories, developers now can face two major challenges: selection of an API amidst multiple 
choices and then learning how to properly use it~\cite{Uddin-OpinerReviewAlgo-ASE2017,Uddin-SurveyOpinion-TSE2019}. 
Both tasks can be facilitated by the official API resources. Unfortunately, 
official API documentation can be incomplete, obsolete and incorrect~\cite{Robillard-FieldStudyAPILearningObstacles-SpringerEmpirical2011a,
Robillard-APIsHardtoLearn-IEEESoftware2009a,Uddin-HowAPIDocumentationFails-IEEESW2015}, 
which often leaves developers no choice but to look for alternative documentation and knowledge sharing resources~\cite{Uddin-SurveyOpinion-TSE2019,Ponzanelli-PrompterRecommender-EMSE2014}. 

The advent and proliferation of online developer forums
has opened up an interesting avenue for developers to look for solutions of
their development tasks in the forum posts~\cite{Barzillay-StackOverflow-Springer2013,Treude-APIInsight-ICSE2016}. 
Among the numerous online forums, Stack Overflow (SO) is a large online community where millions of developers ask and answer questions about their programming needs.  
Developers post questions in SO about their different technical topics, such as selection, usage, and troubleshooting of APIs. Volunteers answer those questions, 
or make comments, as do participants in other social forums. To date, there are around 120 million posts, out 
of which 48 million are questions/answers, and the rest (72 million) are comments. Around 11 million users visit SO and 
add 9K new questions to the site each day~\cite{website:stackexchange-sites}.

The popularity and growing influence of SO has motivated a number of recent research efforts to produce API documentation automatically 
from SO contents, such as adding code examples and interesting textual contents about a Java API type (e.g., a class) 
in the Javadocs~\cite{Subramanian-LiveAPIDocumentation-ICSE2014,Treude-APIInsight-ICSE2016}, 
recommending usage examples within a given IDE~\cite{Ponzanelli-PrompterRecommender-EMSE2014}, summarizing API reviews (i.e., opinions with positive and negative sentiments) to 
assist in API selection~\cite{Uddin-OpinerReviewAlgo-ASE2017}, and so on. In our previous surveys of 178 developers, we find that developers consider the combination of code examples and API reviews 
as a form of API documentation~\cite{Uddin-SurveyOpinion-TSE2019}. In fact, the developers consider such a combination as more valuable 
than official API documentation, when the official resources can be lacking~\cite{Uddin-SurveyOpinion-TSE2019}. We are aware of 
no previous research that attempts to automatically produce API documentation from SO by combining both code examples and reviews.     
  
In this paper, we propose a new documentation format for APIs that we can generate automatically by mining SO. The format considers both  
code example of an API and relevant reviews about the code example from other developers in the forum posts. We present two novel documentation algorithms based on the 
code examples and reviews. The first algorithm is called \bf{Statistical Documentation} which offers visualized usage and review statistics about code examples of an API. 
The second algorithm in \bf{Concept-Based Documentation} which clusters usage scenarios API that are conceptually similar, 
e.g., one scenario consisting of creating an HTTP connection and another sending messages over the HTTP connection. Using the two algorithms, we automatically 
produce the documentation of an API by mining SO. We deploy those documentation in Opiner~\cite{Uddin-OpinerReviewAlgo-ASE2017}. Opiner was previously developed as 
 an online prototype engine to summarize reviews about an API from online developer forums. The overarching goal of Opiner is to 
 become a one stop resource for crowd-sourced API documentation. In this paper, we have extended Opiner to also include our mined usage documentation of the APIs. 
 Opiner is hosted at: \url{http://opiner.polymtl.ca}.     

\begin{figure}[t]
\centering
\vspace{-6mm}
\hspace*{-.8cm}%
\includegraphics[scale=.7]{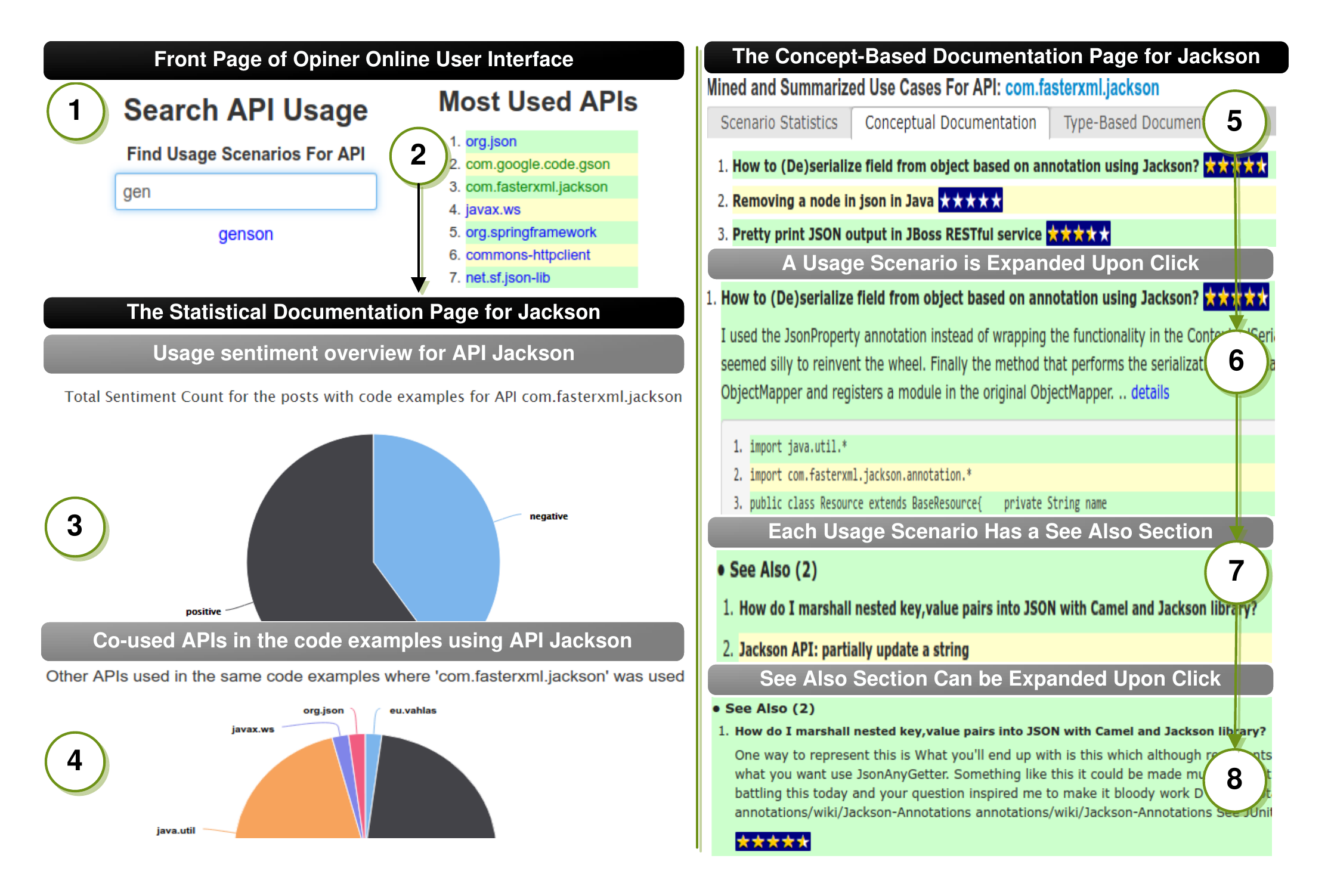}
\hspace*{-1.0cm}%
\vspace{-4mm}
\caption{Screenshots of Opiner online website with the deployed our two novel API usage scenario documentation algorithms. While Opiner online website 
offers other features, the screenshots here only show the extensions of Opiner that were implemented as part of the new contributions presented in this paper}
\label{fig:opiner-motivation-solution}
\vspace{-4mm}
\end{figure}
%
%
%
In \fig\ref{fig:opiner-motivation-solution},
we show screenshots of Opiner usage documentation engine. The Opiner online web-site 
currently indexes the mined and documented usage scenarios of the APIs from a total of 3048 threads SO tagged as `Java+JSON'. This dataset was previously 
used to mine and summarize reviews about diverse Java APIs~\cite{Uddin-OpinerReviewAlgo-ASE2017,Uddin-OpinerReviewToolDemo-ASE2017}. As such, we expect to see 
code examples discussing about Java APIs for JSON parsing 
in the posts. A developer can search for the
usage documentation of an API by searching its name in Opiner - see \circled{1} under the front page of Opiner in \fig\ref{fig:opiner-motivation-solution}. The front page
also shows the APIs with the most number of usage scenarios. As shown in \circled{2}, one of the most used APIs for JSON parsing in Java is Jackson. 
The circles \circled{3} and \circled{4} show some metrics that we developed to produce and visualize the Statistical Documentation.
The circle \circled{3} shows the overall distribution of positive and negative
opinions in the forum posts where the code examples of the API Jackson were found. The circle \circled{4} shows that the API javax.ws (blue pie) is
frequently used alongside the Jackson API in the same code examples. The javax.ws API is an official Java package that is used to create RESTful services, where JSON is the primary 
medium of communication.

The right part of \fig\ref{fig:opiner-motivation-solution} shows screenshots of the concept-based documentation of Jackson in Opiner. 
In Opiner, each concept consists of
one or more similar API usage scenarios. 
Each usage scenario of an API consists of a code example, a textual description of the underlying task addressed by the code example, and reviews (i.e., opinionated sentences with positive/ negative sentiments) of the 
code examples as found in the comments to the post where the code example is found. Each concept is titled as the title of the most
recently posted usage scenario. In  circle \circled{5}, we
show the most recent three concepts for API Jackson. The concepts are sorted by time of their most recent usage
scenarios. The most recent concept is placed at the top of all concepts. Upon
clicking on a concept title, we can see details of the most recent scenario in the concept as shown in circle \circled{6}. Each concept is provided a star
rating as the overall sentiments towards all the usage scenarios that are grouped under the concept (see circle \circled{6}).
 Other relevant usage scenarios of the concept are grouped under a
`See Also' (see circle \circled{7}). Each usage scenario under the
`See Also' can be further explored (see circle \circled{8}). Each usage scenario is
linked to the corresponding post in SO where the code example was
found (by clicking the \it{details} word after the description text of a scenario as shown in \circled{6}).

We evaluated the 
usefulness of the proposed two documentation algorithms over the traditional type-based documentation approach~\cite{Shull-InvestigatingReadingTechniquesForOOFramework-TSE2000}. 
In a type-based documentation, we adopt a Javadoc-style by clustering all the usage scenarios of an API type (e.g., a class) under the type name in Opiner website. 
Previously, Subramanian et al.~\cite{Subramanian-LiveAPIDocumentation-ICSE2014} also promoted similar documentation format for Javadocs by automatically 
mining all the code examples of an API from SO. Given that each usage scenario in our concept-based documentation also contains reviews and textual task 
description of a code example, we added all such information to each code example in our Type-based documentation. We then recruited 29 developers (18 professional) 
and asked them to compare the three documentation types (i.e., Statistics, Concept-Based and Type-Based) 
along four development scenarios (e.g., API selection, documentation) as originally used in~\cite{Uddin-OpinerReviewAlgo-ASE2017}. 
The participants preferred our proposed two algorithms over type-based documentation in all the development scenarios. 

We conducted a second user study 
using 31 developers to evaluate the effectiveness of the produced documentation in Opiner to complete coding tasks. Each participant 
completed four coding tasks using Opiner documentation, official Javadocs, SO, and everything (i.e., including search engine). 
The participants, on average, wrote more correct code, in the least amount time, and using the least effort while using Opiner compared to the 
other documentation resources. In a subsequent survey, more than 80\% participants preferred the Opiner documentation over existing SO 
posts. More than 85\% of participants asked the Opiner documentation platform to be 
 integrated into the formal API documentation to complement and improve the API official documentation.


\begin{table}[t]
  \centering
  \caption{Contributions and Research Advances Made in our Paper}
    \begin{tabular}{lp{3.5cm}p{8.6cm}}\toprule
    \multicolumn{1}{l}{\textbf{Contribution}} & \multicolumn{1}{l}{\textbf{Summary}} & \multicolumn{1}{l}{\textbf{Research Advancement}} \\
    \midrule
    Algorithms &  We propose two novel algorithms to automatically document API usage scenarios from online developer forum: Statistical and Concept-based. & 
    Previous related research focused 
    mainly on linking code example or interesting insights directly to the 
    Javadoc of an API type~\cite{Subramanian-LiveAPIDocumentation-ICSE2014,Treude-APIInsight-ICSE2016}, 
    or to complement API official documentation using SO contents~\cite{Wang-APIsUsageObstacles-MSR2013,Kavaler-APIsUsedinAndroidMarket-SOCINFO2013,Souza-CookbookAPI-BSSE2014,Sunshine-APIProtocolUsability-ICPC2015,Mastrangelo-JavaUnsafeAPIs-OOPSLA2015,YeDeheng-ExtractAPIMentions-ICSME2016,Campos-SearchSORecommend-JSS2016,Campos-SearchSOPostAPIBug-CASCON2016,Azad-GenerateAPICallrules-TOSEM2017,Ahsanuzzaman-ClassifySOPost-SANER2018,Wang2013Detecting,Dagenais-DeveloperDocumentation-FSE2010a,Parnin-MeasuringAPIDocumentationWeb-Web2SE2011,Parnin2012,Jiau-FacingInequalityCrowdSourcedDocumentation-SENOTE2012,Campbell-DeficientDocumentationDetection-MSR2013,Treude-DocumentationInsightsSO-ICSE2016,Delfim-RedocummentingAPIsCrowdKnowledge-JournalBrazilian2016,LiXing-LeveragingOfficialContentSoftwareDocumentation-TSC2018,LiSun-LearningToAnswerProgrammingQuestions-JIS2018}. 
    Our algorithms offer directions to design innovative algorithms to complement and improve API official documentation. \\
    \midrule
    Techniques &  We implemented and deployed the algorithms in our tool, Opiner~\cite{Uddin-OpinerReviewToolDemo-ASE2017}.  
    &   
    We are aware of no tool that can offer search and documentation features of API usage scenarios 
    automatically collected from developer forums. The underlying documentation framework in Opiner can be further extended 
    with new API usage documentation algorithms.\\
    \cmidrule{2-3}
     &  
    We conducted three user studies to demonstrate the effectiveness of the proposed usage documentation algorithms over traditional API documentation approach and resources. 
     & 
    The positive reception of our proposed API documentation formats based on the two algorithms opens up a new research area in software engineering 
    to design innovative techniques and tools by harnessing knowledge shared in online crowd-sourced forums. 
    As we noted in \secs\ref{sec:introduction} and \ref{sec:related-work}, existing 
    research~\cite{Subramanian-LiveAPIDocumentation-ICSE2014,Treude-APIInsight-ICSE2016,Souza-CookbookAPI-BSSE2014,Sunshine-APIProtocolUsability-ICPC2015,Campos-SearchSORecommend-JSS2016,Campbell-DeficientDocumentationDetection-MSR2013,Treude-DocumentationInsightsSO-ICSE2016,Delfim-RedocummentingAPIsCrowdKnowledge-JournalBrazilian2016,LiXing-LeveragingOfficialContentSoftwareDocumentation-TSC2018,LiSun-LearningToAnswerProgrammingQuestions-JIS2018} mostly focused on complementing the traditional official documentation. 
    \\
    \bottomrule
    \end{tabular}%
  \label{tab:researchAdvancement}%
\end{table}%

In summary, we advance the state of the art by presenting two novel algorithms to automatically document API usage scenarios from online developer forums with each deployed in 
 an online API documentation prototype tool Opiner. We demonstrate the effectiveness of the algorithms 
 and the tool to assist developers in their diverse development tasks using three user studies. In \tbl\ref{tab:researchAdvancement}, 
 we outline the major contributions of this paper.

%% file: researchContext.tex
\section{Background}\label{sec:background}
 This research borrows concepts and techniques from software engineering and opinion analysis. In this section, we present the major concepts and techniques upon which this study is founded.
 \subsection{API}
In this paper, we investigated our API usage documentation techniques for both open-source and official Java APIs. As such, we analyzed
SO posts tagged as ``Java'' where Java APIs are mostly discussed. However, the analysis
and the techniques developed can be applicable for any API.
In particular, we adopt the definition of an API pioneered by Martin Fowler. An API is a ``set
of rules and specifications that a software program can follow to access and make
 use of the services and
resources provided by its one or more modules''~\cite{website:wikipedia-api}.
 An API is identified by a name. An API
consists of one or more modules.
Each module can have one or more source code packages. Each package can have one or
more code elements, such
as classes, methods, etc.
For the Java official APIs available through the
Java SDKs, we consider an official Java package as an API. Similar
format is adopted in the Java official documentation (e.g., the \tt{java.time}
package is denoted as the Java date APIs in the
new JavaSE official tutorial~\cite{website:oracle-javadateapi}).

As shown in \fig\ref{fig:HowAPIsDiscussed}, this is also how APIs are discussed and mentioned in SO. For example, 
there are three open source Java APIs mentioned in the textual contents in \fig\ref{fig:HowAPIsDiscussed}: Jackson, Google Gson, and org.json. 
In the code example, two packages from the official Java SDK are used along with Gson: java.util and java.lang.

An API is normally designed to support specific development needs. Each need can be implemented as a functionality in the API. Each functionality 
is denoted as `feature'~\cite{Robillard-APIProperty-IEEETSE2012}. For example, the Gson API is developed to support the processing and manipulation of JSON-based inputs in Java. One feature of the Gson 
API is the conversion of JSONArray into a Java Object. As shown in \fig\ref{fig:HowAPIsDiscussed}, this can be addressed by using two methods 
from the two classes of the Gson API: getType(\ldots) and fromJson(\ldots) from the classes TypeToken and Gson, respectively.

\subsection{Opinion}
Bing Liu, in his book~\cite{liu-sentimentanalysis-handbookchapter-2016}, defines opinion
as: ``An opinion is a quintuple $<$$e_i$, $a_{ij}$, $s_{ijkl}$,
$h_k$, $t_l$$>$, where $e_i$ is the name of the entity, $a_{ij}$ is an aspect of
$e_i$, $s_{ijkl}$ is the sentiment on aspect $a_{ij}$ of entity $e_i$, $h_k$ is
the opinion holder, and $t_l$ is the time when the opinion is expressed by
$h_k$". The sentiment $s_{ijkl}$ is positive or negative. Both entity ($e_i$) and
aspect ($a_{ij}$) represent the opinion target. An aspect about an entity can be about a property or a feature supported by the entity. 

\begin{figure}[t]
  \centering
  \vspace{-6mm}
   \hspace*{-.4cm}%
  \includegraphics[scale=.65]{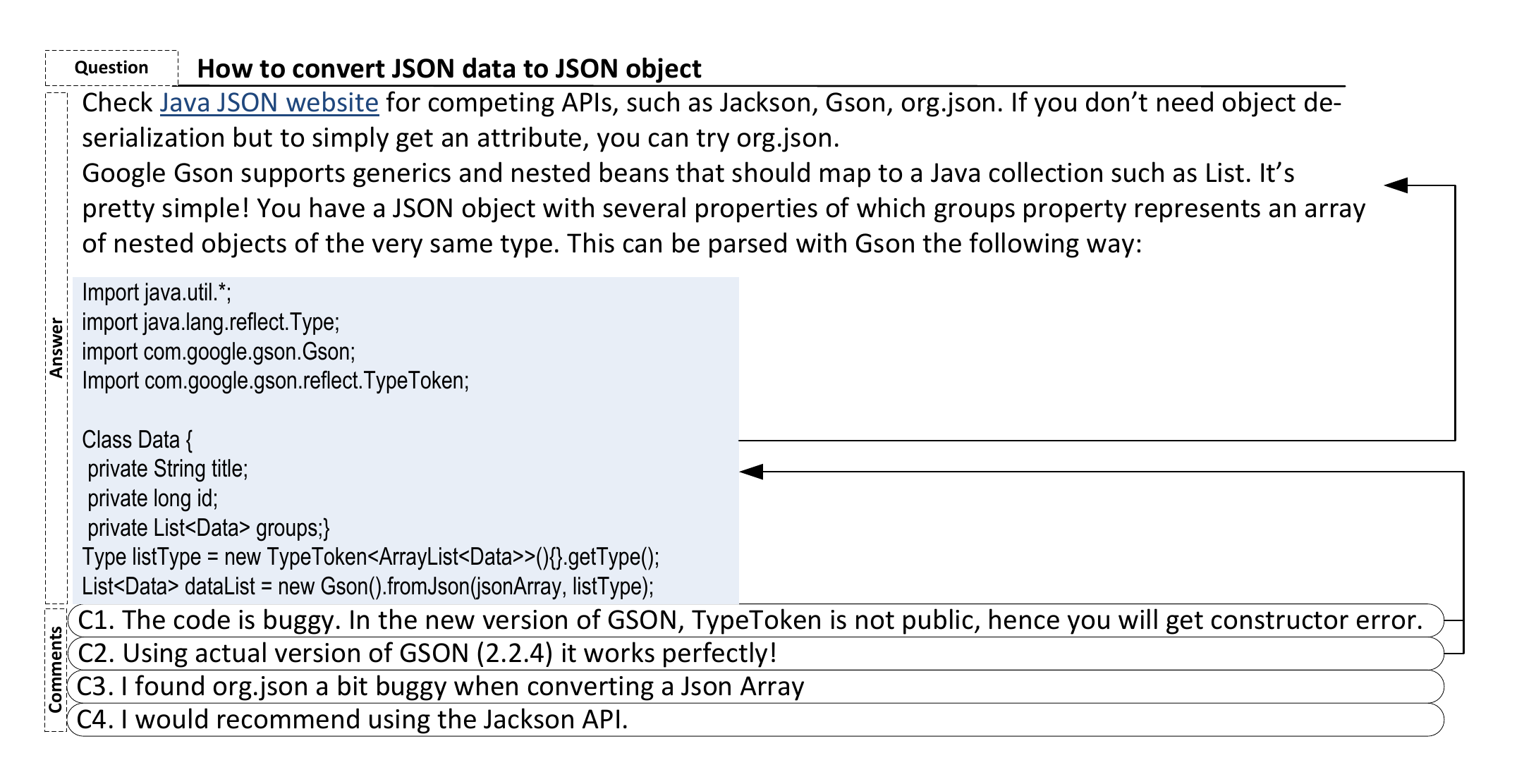}
  \vspace{-4mm}
  \caption{How APIs are discussed in SO}

  \label{fig:HowAPIsDiscussed}
\vspace{-4mm}
\end{figure}

For example, in \fig\ref{fig:HowAPIsDiscussed} the first comment (C1) has two sentences. The first sentence is `The code is buggy'. This a negative opinion 
about the the bug aspect of the provided code example.   

\subsection{API Usage Scenario}\label{sec:apiUsageScenario}
In this paper, we produce API documentation by combining code examples of an API with the relevant reviews towards the code examples. We use the notion `API Usage Scenario' which is a composite of three items: a code example associated to an API, a textual description of the underlying task addressed by 
the code example, and a set of reviews (i.e., opinions with positive and negative sentiments) towards the code example as provided in the comments to the post where the code example is found. 
For example, from \fig\ref{fig:HowAPIsDiscussed}, we can produce an API usage scenario based on the code snippet as follows:
\begin{inparaenum}
\item The code snippet is provided to complete a development task involving Java to convert JSON data to Java object using the Google Gson API.
\item A textual task description of the task by identifying relevant sentences, such as those immediately before the code snippet. 
\item The reviews in comments C1 and C2 that are relevant to code snippet.
\end{inparaenum} 

Our decision to use API usage scenarios instead of simply code examples is influenced by seminal research of Carroll et al.~\cite{Carroll-MinimalManual-JournalHCI1987a} and Shull et al.~\cite{Shull-InvestigatingReadingTechniquesForOOFramework-TSE2000}.
Carroll et al.~\cite{Carroll-MinimalManual-JournalHCI1987a} proposed `minimal manual' for technical documents by designing the documentation around specific tasks. 
In subsequent study, Shull et al.~\cite{Shull-InvestigatingReadingTechniquesForOOFramework-TSE2000} find that such a task-based documentation format is more useful than a traditional hierarchical 
documentation format~\cite{Shull-InvestigatingReadingTechniquesForOOFramework-TSE2000}.   
In our API usage scenarios, each scenario corresponds to a specific development task.   

Our decision to utilize reviews from comments is based on our previous findings from surveys of 178 software developers from SO and GitHub~\cite{Uddin-SurveyOpinion-TSE2019}. 
We find that developers consider the combination of a code example in an answer post and the reviews about it from other developers in the comments as a form of API documentation. We 
also find that developers consider such a combination more valuable than API official documentation, because the reviews are offered by experts and are based on 
their real-world experience on the API usage. The usefulness of comments is confirmed with empirical evidence by two recent studies as well, published at the same year 
of our surveys (i.e., 2019). Ren et al.~\cite{Ren-DiscoverControversialDiscussions-ASE2019} 
exploited comments to identify `controversial' answers, i.e., answers that may be potentially incorrect. They find that in those `controversial' cases, comments 
are useful to offer a more accurate usage experience of the API. In a separate study, Zhang et al.~\cite{Zhang-ReadingAnswerSONotEnough-TSE2019} manually analyzed a statistically 
significant sample of all SO comments and found that more than 75\% of the comments are useful. Indeed, the number of comments is much more than the number of 
answers in SO (72 million vs 29 million as of 2020). Zhang et al.~\cite{Zhang-ReadingAnswerSONotEnough-TSE2019} further emphasized: \emt{The amount
of information in comments cannot be neglected, with  23\% of the answers having a commenting-thread that is longer than their actual answer.} These positive findings 
from Zhang et al.~\cite{Zhang-ReadingAnswerSONotEnough-TSE2019} highlight that most of the comments in SO are informative and not noisy, and thus could be 
used to assist developers.

\subsection{Javadoc}
Our API documentation framework in Opiner currently supports our two proposed algorithms and a Javadoc-style presentation of the mined API usage scenarios. 
In the Javadoc of an API, individual pages are created to document each type of the API. A type for a Java API can be a class, annotation, or an interface. 
The Javadoc-style presentation in Opiner is called Type-based documentation, because we cluster all usage scenarios associated to an API type under the type. 
For Java APIs, Javadocs are one of the most commonly known and used documentation formats - as noted in a number of old and recent previous research~\cite{Ponzanelli-PrompterRecommender-EMSE2014,Shull-InvestigatingReadingTechniquesForOOFramework-TSE2000}. 
As such, previous research on automatic API documentation efforts have proposed to augment the Javadocs of an API type with code examples and 
relevant insights (e.g., specific conditions of usage) from SO~\cite{Subramanian-LiveAPIDocumentation-ICSE2014,Treude-APIInsight-ICSE2016}. 
However, previous research finds  that a Javadoc-type hierarchical documentation format is not a useful presentation format~\cite{Uddin-HowAPIDocumentationFails-IEEESW2015}. 
We thus innovate by proposing two novel API documentation algorithms, which are different from the type-based documentation format. 

%% file: summarizationAlgos.tex
\section{The API Documentation Framework}\label{sec:framework}
\begin{figure}[t]
\hspace*{-.7cm}%
  \centering \includegraphics[scale=.86]{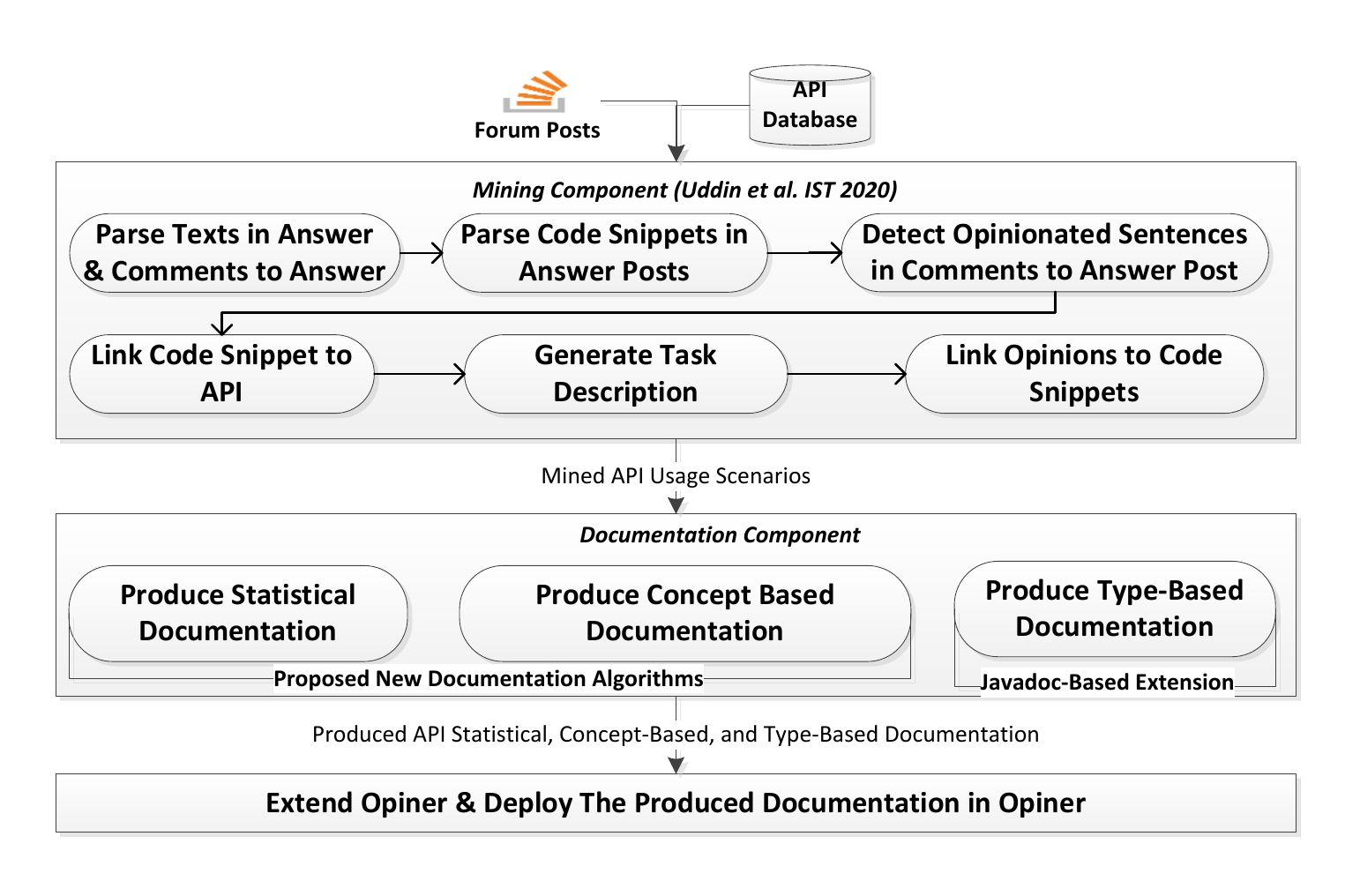}
  \hspace*{-.7cm}%
   \vspace{-4mm}
  \caption{The major components in our proposed crowd-sourced API documentation framework}
  \label{fig:framework}
\end{figure}
The input to our API documentation framework is a list of forum posts and an API database. The final outputs are API documentation 
based on the input. The framework consists of two major components (see \fig\ref{fig:framework}): 
\begin{enumerate}
\item Mining Component. Takes as input the forum posts and an API database. The output is a list of API usage scenarios. Each API scenario consists of a code snippet, textual description, and reviews towards the snippet. 
\item Documentation Component. Takes as input the mined usage scenarios of an API and produces three types of documentation. The first two are based on our two proposed documentation algorithms (statistical and concept-based). 
The third is type-based documentation, which is developed as an adaptation of Javadoc.  
\end{enumerate} Our API database consists of \begin{inparaenum}
\item all the Java APIs collected from
the online Maven Central repository~\cite{website:maven-central},\footnote{We consider a binary file (e.g., jar) of a project from Maven as an API} 
and
\item  all the Java official APIs from JavaSE (6-8)
and EE (6 and 7). 
\end{inparaenum} In \tbl\ref{tbl:apidb}, we 
show the summary statistics of our API database.
All of the APIs (11,576) hosted in Maven central are for Java. 
We picked the
Maven Central because it is the primary source for hosting
Java APIs with over 70 million downloads every
week~\cite{website:mavencentral-blog}.  
\begin{table}[t]
\caption{Descriptive statistics of the Java APIs in the database}
\centering
\begin{tabular}{llll|rrr}\toprule
\textbf{API} & \textbf{Module} & \textbf{Version} & \textbf{Link} & \textbf{\#Avg Links/API}
& \textbf{\#Avg Modules/API} & \textbf{\#Avg Versions/Module} \\ \midrule
12,451 & 144,264 & 603,534 & 72,589 & 5.8 & 11.6 & 4.2 \\ 
\bottomrule
\end{tabular}
\label{tbl:apidb}
\end{table}
We describe the components below.

\subsection{Mining Component}\label{sec:mining-component}
\rev{Given as input a forum post, first we preprocess the post contents and then we mine API usage scenarios from the parsed contents. 
The techniques supporting both steps are previously published in Uddin et al.~\cite{Uddin-MiningAPIUsageScenarios-IST2020}. We thus briefly describe the steps below and leave the 
details in \cite{Uddin-MiningAPIUsageScenarios-IST2020}.} 
 
Given as input an answer to a question in SO, we divide its contents using three parts:
\begin{inparaenum}[(1)]
\item Code snippet in the answer. We detect code snippets as the tokens wrapped with the
 $<$code$>$ tag.
 \item Textual contents in answer, and 
 \item Textual contents in the comments to the answer. 
\end{inparaenum} The textual contents are tokenized into sentences. Opinionated sentences are detected 
as sentences having positive or negative polarity. To detect opinionated sentences, we use the OpinerDSO algorithm~\cite{Uddin-OpinionValue-TSE2019} 
which offered performance comparable to state-of-the-art sentiment detection tools for software engineering, e.g., Senti4SD~\cite{Calefato-Senti4SD-EMSE2017}. 
Nevertheless, the Opiner framework is flexible to replace the OpinerDSO by any other sentiment 
detector. During the parsing of code snippets, 
we discard non-code and non-Java snippets (e.g., XML, Javascript) using Language-specific 
naming conventions (similar to Dagenais and Robillard~\cite{Dagenais-RecoDocPaper-ICSE2012a}). 
We parse a valid code example to identify API elements (types, methods, interfaces). We consult our API database to infer FQN (Fully Qualified Name) of the API elements. 
Given as input a parse code example and the textual contents in the post where code example is found, we produce an API usage scenario using three algorithms as follows. 

\rev{First, we heuristically link a code snippet to an API name mentioned in the textual contents
of the forum post where the code snippet is discussed by consulting
the textual contents and API elements found in the code snippet.
For example, in \fig\ref{fig:HowAPIsDiscussed}, the code snippet is linked to
the Google Gson API. State-of-the-art algorithm like
Baker~\cite{Subramanian-LiveAPIDocumentation-ICSE2014} is not designed to link a
code example to an API name mentioned in the textual contents of a forum post.
For example, for the code snippet in \fig\ref{fig:HowAPIsDiscussed},
Baker~\cite{Subramanian-LiveAPIDocumentation-ICSE2014} links it to three APIs
(java.util, java.lang, and Google GSON). However, the code snippet is provided
to explain the conversion of JSON data to JSON object using the GSON API, as
mentioned in the textual contents. The code snippet also has the most number of API classes and
methods matched with those found in the Gson API. Second, we produce a textual description of
the underlying task addressed by the code snippet. The algorithm does this by
picking sentences in the textual contents of the forum post where the code
snippet is discussed and where the API linked to the code snippet is referred
to. For example, in \fig\ref{fig:HowAPIsDiscussed}, the following sentence is
picked into (among others) the description, ``Google Gson supports generics and
nested beans \ldots'', but not this sentence ``if you don't need object
deserialization, \ldots you can try org.json''. The description is produced by
combining beam search~\cite{Reddy-BeamSearch-CMU1977} with TextRank
algorithm~\cite{Mihalcea-Textrank-EMNLP2004}. Third, we associate positive and negative opinions  relevant to the code example
by analyzing the comments to the post. The algorithm does this by looking for
references in the comments to the API that is linked to the code snippet. For
example, in  \fig\ref{fig:HowAPIsDiscussed}, all opinionated sentences from
comment C1 and C2 are linked to the code snippet. Each algorithm shows a
precision and recall of over 0.8 on multiple evaluation settings. Each algorithm
outperforms the state-of-the-art baselines (e.g.,
Baker~\cite{Subramanian-LiveAPIDocumentation-ICSE2014}). For further 
details we refer to the paper~\cite{Uddin-MiningAPIUsageScenarios-IST2020}.}

\subsection{Documentation Component}\label{sec:documentation-component}
%

Given as input the mined usage scenarios of an API, we produce three types of documentation: two using our two proposed novel documentation 
algorithms (statistical, concept-based) and the third 
based on an adaption of Javadoc presentation format. We discuss the three documentation types below.


\subsection*{Algorithm 1. Statistical Documentation}\label{sec:statistical-summary}
We produce statistical documentation of the mined API usage scenarios to offer a visualized overview of the usage of an API based on the mined usage scenarios. 
This documentation thus can complement the front/introductory page of a API documentation by offering visualized statistics of the API usage. In addition, 
this type of documentation can also offer a quick overview of the underlying quality of the code example (described below).  

To complement the front page of API documentation, we produce three types of visualizations: 
\begin{enumerate}
\item {Sentiment Overview.} The overall sentiments in the reactions to the
usage scenarios of the API, and
\item{Co-Used APIs.} The usage of other APIs in the same code examples of the
API.
\item {Co-Used API Types.} How the various types (e.g., classes) of an API are often used together.
\end{enumerate}  
In addition, we provide an overview of the quality of each API usage scenario as follows:
\begin{enumerate}[start=4]
\item {Star Rating of API Usage Scenario.} Following previous research and adoption of 5-star ratings in online product reviews~\cite{Liu-SentimentAnalysisOpinionMining-MorganClaypool2012} 
and API review summarization~\cite{Uddin-OpinerReviewAlgo-ASE2017,Uddin-OpinerReviewToolDemo-ASE2017}, we show the overall rating of each API usage scenario 
by analyzing the positive and negative opinions related to a code example.
\item {Star Rating of API Type.} The overall star-rating of an API type based on the usage scenarios of the API where the type was used.
\end{enumerate} We describe the approaches below.  

\begin{inparaenum}
\nd\item\ul{Sentiment Overview.} In our previous work on API
review summarization~\cite{Uddin-OpinerReviewAlgo-ASE2017}, we observed that
developers reported to have increased awareness about an API while 
seeing the visualized overview of the API reviews from developer forums. We thus offer two types of overview of the quality of all code examples of an API.
The first is a simple pie chart by showing the overall counts of all the positive and negative reactions to the usage scenarios linked to an API. 
We do this aggregating all the positive and negative opinions across all the mined usage scenarios of an API. 
The second is a time-series of the aggregated sentiments. We do this as follows. 
\begin{inparaenum}
\item We find the first and last month-year of creation dates among the usage scenarios of an API. The creation date of a usage scenario is the time when the post containing the 
usage scenario was created. 
\item For each month-year, we create two bins: positive and negative. The posiive bin contains the count of all positive opinions the posts created in that month-year received due to the 
code examples discussed in those posts and included in our API usage scenarios. 
\item We create two time-series one for positive and another for negative polarities. Each time-series chart has an X-axis as month-year and y-axis as the counts of positive/negative 
polarity for that month-year. 
\end{inparaenum} While the pie-chart offers overall sentiments of developers towards the usage scenarios of an API, the time-series chart shows 
how the sentiments changed over time. For example, an API may have high positive reviews when it was first created, but it started get more negative reviews over 
time due to obsolete code examples. 


\nd\item\ul{Co-Used APIs.} For a given API, we show how many other APIs were
used in the code examples where the API was used. Because each API offers features focusing on specific development scenarios, 
such insights can be helpful to know which other APIs besides the given API a developer needs to learn to be able to properly utilize the API for an end-to-end development solution. 
We create a pie-based presentation as follows:
\begin{inparaenum}[(a)]
\item Identify all the other APIs discussed in the same code example. We did
this by looking at imports and by matching the different types in the code
example against all the APIs in the mention candidate lists associated with the
code example.
\item Take a distinct count of API for each example. Thus the bigger a
pie in the chart, the more number of code examples were found where the two APIs
were used together.
\end{inparaenum} 

\nd\item\ul{Co-Used API Types.} For a given type (e.g., a class) of an API, we show how frequently other
types of the same API were used together. We create a pie-chart
that shows how frequently other types from the 
same API were co-used in the same code example where the given API type was used.

\nd\item\ul{Star Rating of API Usage Scenario.} We compute the overall five star rating of a mined API usage scenario 
by taking into account all the positive and negative reactions towards the code example in the scenario. The computation is 
adopted from the star rating schemes we proposed
previously~\cite{Uddin-OpinerReviewAlgo-ASE2017,Uddin-OpinerReviewToolDemo-ASE2017}.
For every concept of an API, we collect all the positive and negative opinions and compute the star rating as:
\begin{equation}\label{eq:star-rating}
\tr{Star Rating} = \frac{| Positives |\times 5 }{ | Positives | + | Negatives |}
\end{equation}

\nd\item\ul{Star Rating of API Type.} We compute the overall five star rating (using \eq\ref{eq:star-rating}) of each type
by taking into account all the positive and negative reactions towards the usage
scenarios grouped under the type. 
\end{inparaenum}


\subsection*{Algorithm 2. Concept Based Documentation}\label{sec:concept-summary}
\rev{The input to the concept-based documentation algorithm is a list of all usage scenarios associated to an API. The 
output is a list of concepts, where each concept contains a list of API usage scenarios that are \it{similar} to each other based on the development tasks 
implemented. We propose concept-based documentation to present the mined API usage scenarios by grouping the scenarios around conceptually similar tasks. 
A \it{concept} in our documentation algorithm is a \it{cluster} of API usage scenarios that 
offer \it{similar} features or that are \it{situationally relevant}. Two API usage scenarios are similar, if the code examples 
use the same API elements (e.g., classes) and they have similar inputs and outputs. For example, if two code examples 
using the GSON API offer conversion of JSON objects to Java objects, they have similar inputs (e.g., JSON object) and similar outputs (e.g., Java objects). 
As such, we cluster the two code examples into one concept. Two API usage scenarios are situationally relevant, if the code examples 
have similar inputs, but they produce different outputs and vice versa. For example, two code examples using the GSON are situationally relevant if 
both take as input a JSON object, but one offers conversion of the JSON object to Java objects and another to XML objects.}

\begin{figure*}[t]
\centering
\includegraphics[scale=.7]{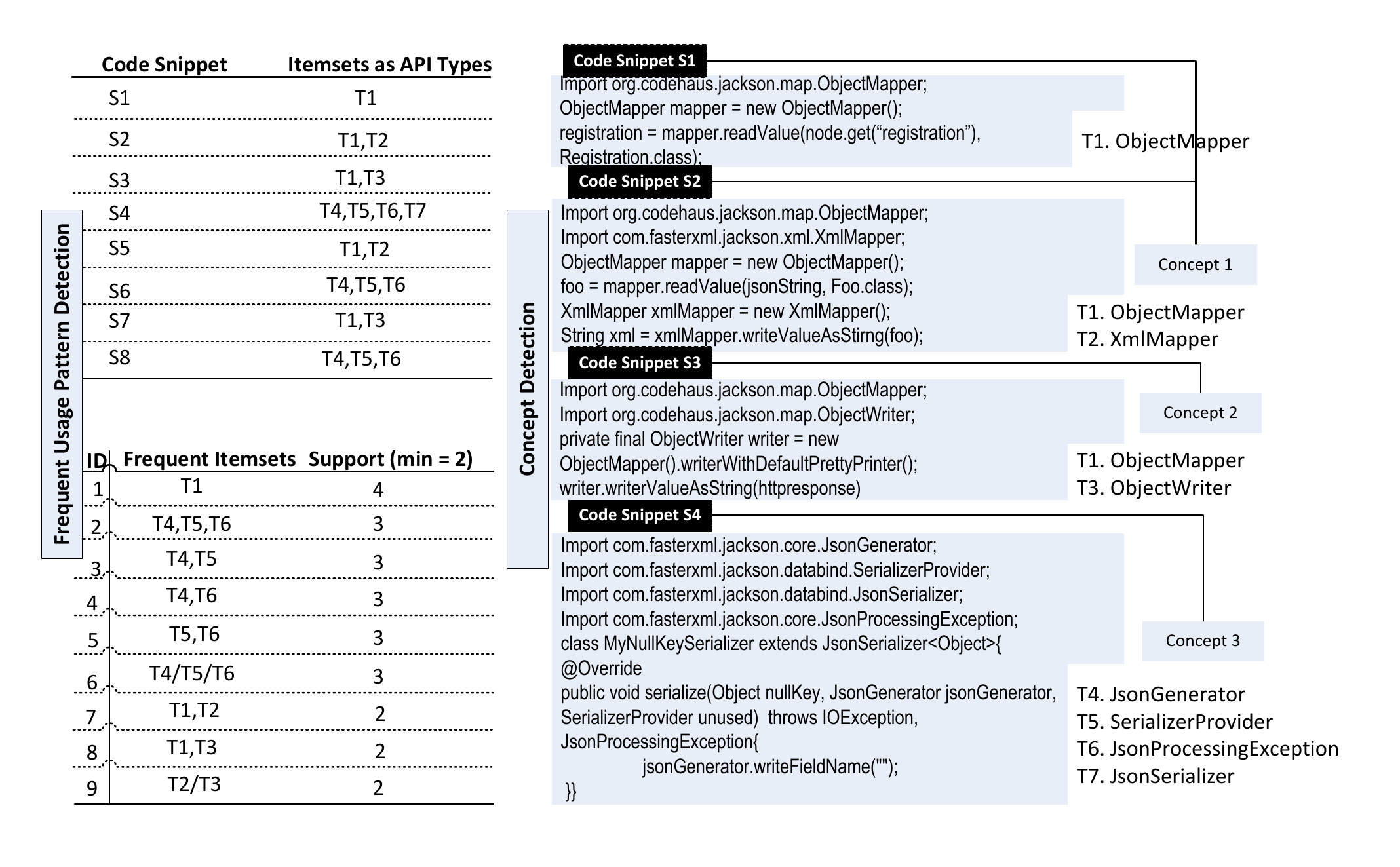}
\hspace*{-2cm}%
\caption{Examples on how usage patterns and concepts are detected in the concept-based documentation of an API}
\label{fig:concept-summary-example}
\end{figure*}

Given that the code examples in a concept should have similar inputs, the code examples should then use similar API elements. That means the code examples 
in a concept should exhibit similar usage pattern involving similar API elements. Therefore, 
given as input all the usage scenarios of an API, we first identify \it{frequent itemsets} as types (e.g., classes) of an API 
that are found to be used frequently together. A set of code examples showing similar usage patterns can be similar, if 
they are clone of each other or if they conceptually relevant (e.g., similar input or similar output). Therefore, our approach has two major steps:
\begin{enumerate}
  \item {Detect Usage Patterns.} We detect API types as itsemsets that are frequently used together in the scenarios. We assign usage scenarios to the patterns.
  \item {Detect Concepts.} We create clusters of API usage scenarios in the detected usage patterns that are similar to each other based on inputs and/or outputs. Each cluster is denoted as a `concept'.   
\end{enumerate}
%
We provide visualized examples of the two steps in \fig\ref{fig:concept-summary-example} and describe the steps below.

\nd\ul{Step 1. Detect Usage Patterns.} Given that similar code examples in our concepts should use similar API elements, we 
first need to identify \it{patterns} of API elements in the mined usage scenarios of an API that are frequently used together. 
To detect usage patterns, we use frequent itemset mining~\cite{Pasquier-FrequentClosedMining-ICDT1999}.
Frequent itsemset mining has been used in
summarizing product features~\cite{Hu-MiningSummarizingCustomerReviews-KDD2004}
and to find useful usage patterns from software
repositories~\cite{Bettenburg:2013:DSS:2486788.2486960,ZimmermannZeller-eRose-TSE2006a}. 

Using frequent 
itemset mining, we cluster 
code examples that use the same types of a given API, even that the code example can use more than one API. For example, if a code example is 
provided to convert JSON string to Java object using the GSON API, the JSON string can come from a file, from a web service, or from any source. 
The code example thus can show usage of the reading of the JSON string from those sources using APIs other than the GSON API. A clone detection 
technique may consider such code examples as not clones, because one code example may be using the java.util API to read a JSON string from a file while another 
may be using the HttpClient API to read the JSON string from web server. However, both code examples are using the same GSON API types. As such 
both will be clustered under a concept using our frequent itemset mining approach.

In \fig\ref{fig:concept-summary-example}, the left 
column shows examples of detecting usage patterns. The right column shows four of the code snippets (S1-S4) used in the examples of left column. 
All the code snippets in \fig\ref{fig:concept-summary-example} are associated to the API Jackson for JSON parsing. 
Given as input all the mined API usage scenarios of an API, our pattern detection involves two steps: \begin{inparaenum}
\item Collection of API types from the code examples, and 
\item Generation of frequent sets of API types. 
\end{inparaenum} 

First, for each code example linked to an API, we collect all the types (class, exception, annotation, etc.) 
  found in the code example that belong to the API of our interest. For example, for the code snippet S1 in  \fig\ref{fig:concept-summary-example}, only one type (\tt{ObjectMapper}) is 
  collected, because this is the only type from Jackson API in S1. For the code snippet S2, we find two types from Jackson API (\tt{ObjectMapper} and \tt{XmlMapper}). 
  Similarly, given we focus on clustering types of given API, which is Jackson in \fig\ref{fig:concept-summary-example}, we ignore types that 
  come from other APis or that are local, such as Registration in S1, foo and String 
  in S2, etc. Registration and foo are ignored, because those are local classes, i.e., those are not provided by the Jackson API. The class `String' is ignored, because it is provided 
  by java.lang package. \rev{Please note that our linking of a code example to
  an API name mentioned in the forum text considers all the code elements from
  all APIs in a code example (method, class, interface, etc.), as we describe
  the linking algorithm in our paper~\cite{Uddin-MiningAPIUsageScenarios-IST2020}. While producing concepts by taking as
  input all code examples linked to an API, we focus on types (class, interface)
  of a given API in the code examples, because we focus on
  clustering code examples implementing similar features of the API and that
  previous studies find that API types are more informative than API methods to
  analyze API features~\cite{Dagenais-DeveloperLearningResources-PhDThesis2012}.}
  
Second, we create a list of itemsets by collecting API types as explained above. 
  Thus for eight code examples in \fig\ref{fig:concept-summary-example} left column, 
  we have eight itemsets.  Second, we apply the Frequent Itemset Mining~\cite{Pasquier-FrequentClosedMining-ICDT1999} on the lists using 
  the FPGrowth C implementation developed by Borget~\cite{Borgelt-FIM-Wiley2012}. The output is a list of frequent itemsets and a support value for each itemset. For example, the second frequent itemset in \fig\ref{fig:concept-summary-example} (ID 2) 
is $\{T4,T5,T6\}$ with support 3 (because it is found in three code snippets S4, S6, S8). Each frequent itemset is considered as a pattern. For example, in \fig\ref{fig:concept-summary-example} we have eight 
patterns that were found in at least two code examples. We assign a code snippet to a pattern by computing the similarity between a code snippet ($S$) and a pattern ($P$)
as follows:
\begin{equation}
\tr{Similarity} = \frac{|\tr{Types}(S)\bigcap
\tr{Types}(P)|}{|\tr{Types}(S)|}
\end{equation} We assign a code example to the pattern with the highest similarity. For example, all the three code examples (S4, S6, S8) are 
assigned to the pattern $\{T4,T5,T6\}$ (ID 2) in \fig\ref{fig:concept-summary-example}. If more than one pattern is found with the maximum similarity value, 
we assign the code snippet to the pattern with the maximum
support. Therefore, the output of this step is a matrix $P\times S$, where $P$ stands for a usage pattern and $S$ stands for a set of usage scenarios assigned to the 
pattern.

More than one API usage scenario can belong to one concept. For example, all the
code examples related to the frequent itemsets (T4, T5), (T4,T6), and (T5,T6)
will be grouped under the super itemset (T4,T5,T6) in
\fig\ref{fig:concept-summary-example}. We do this to ensure that a concept can
contain all the different use cases (i.e., scenarios) that use the API types
under the concept. \rev{The current implementation of concepts in our algorithm
assigns each API usage scenario into one concept only, i.e., it is a `hard'
assignment. Intuitively, an API usage scenario can be similar to API usage
scenarios assigned to more than one concept, e.g., due to situational relevance.
We leave the creation and analysis of such `soft' assignment of API usage
scenarios into concept as our future work.} Given that all such relevant code
examples are grouped under a concept, a developer in our produced API
documentation would only need to look at one or two of the grouped API usage
scenarios in the concept to get a concise but complete insight of the overall
API usage addressed by the concept.
Another way to produce such concepts would be to use closed frequent
itemsets~\cite{Pasquier-FrequentClosedMining-ICDT1999}. However, closed frequent
itemsets only return the super (i.e., closed) itemsets. Therefore, while our
concept detection approach borrows ideas from closed frequent itemset to find
super itemset, we leverage standard frequent itemset mining to also identify all
the frequent itemsets under the closed itemset.

The output from this step is a matrix of patterns vs mined API usage scenarios $P\times S$, where P stands for a pattern and S contains 
a list of code examples belonging to the pattern. The code examples in a pattern of an API 
contain a set of commonly co-used 
types of the API. 

\nd\ul{Step 2. Detect Concepts.}  In this step, we further analyze the 
patterns identified in the previous step to determine whether two or more patterns could be connected with each other. Intuitively, a connection between two patterns can be established 
if code examples in the two patterns share similar inputs/outputs. For example, suppose one pattern contains code examples related to the establishment of an HTTP connection 
using the HttpClient API and another pattern contains code examples related to the sending/receiving or messages over an HTTP connection. The two patterns 
can be connected, because the output (i.e., an established HTTP connection) from pattern 1 (i.e., the code examples in pattern 1) is used as input in pattern 2.
If we find such connected patterns, we group those patterns together into a concept. We detect concepts as follows. 

First, given all code examples under a pattern, we apply clone detection to find code examples that are similar to each other. This helps us 
create intermediate sub-groups of code examples in a pattern, where code examples in a sub-group are clones of each other. We then compare the inputs/outputs 
between two patterns by comparing the inputs/outputs in the sub-groups found in the two patterns. Thus the formation of sub-groups reduces the number of analysis (of inputs/outputs) 
between patterns. This step is important, because otherwise we will be left with an exponential combination of multiple patterns to analyze.     
For clone detection, we use NiCad~\cite{CKRoy-Nicad-ICPC2008}, which is a widely used state of the art clone detection tool in software engineering literature. 
Specifically, we  use NiCad 3, which detects near-miss clones\footnote{Following~\cite{Svajlenko-EvaluatingCloneDetectionTools-ICSME2014}, 
we set a minimum of 60\% similarity and a minimum block size of five lines in the code example. }.
We detect inputs to a code example by analyzing the inputs taken 
by methods in the code example, where the inputs are generated by another method in the code example. We detect outputs from a code example 
as the outputs from the methods, where the outputs are not fed into other method(s) in the same code example.

As a demonstration of the above process, consider \fig\ref{fig:concept-summary-example} right column: The first code snippet (S1) belongs to pattern with ID 1 (left column) and 
the second code snippet (S2) belongs to pattern with ID 7 (left column). S1 uses one type of API Jackson (\tt{ObjectMapper}), S2 uses both 
\tt{ObjectMapper} and \tt{XmlMapper}. Both code snippets use the \tt{readValue} method of the \tt{ObjectMapper} to convert a JSON string 
into a Java class. Code snippet S2 further attempts to produce an XML string out of the generated Java class. It does that 
by using the \tt{XmlMapper} class of the same Jackson API and by taking as input the generated Java class from \tt{ObjectMapper}. 
Therefore, we say Patterns 1 and 7 are situationally relevant, because Pattern 1 always needs to be executed before Pattern 7 above. 
To ensure that we establish the relationship between patterns correctly, we employ program slicing. For example, the third 
code snippet (S3) in \fig\ref{fig:concept-summary-example} right column belongs to pattern with ID 3. S3 also uses the \tt{ObjectMapper} class of the Jackson 
API. However, S3 does not use the \tt{readValue} method of \tt{ObjectMapper}. Instead S3 creates a string out of a \tt{httpresponse} and assigns that to 
another class of the Jackson API (\tt{ObjectWriter}) for pretty printing. We thus do not create an edge between Pattern 1 and Pattern 8.
Similarly, all the code examples (S4, S6, S8) belong to another concept. 

Once a set of patterns are grouped together into a concept, we assign a numeric ID to the concept. 
The output of this step is a matrix $C\times S$, where $C$ stands for a concept and $S$ denotes a list of API usage scenarios assigned to the concept. 
In Opiner web site, we present each concept as a list of four items: 
\{$R, S, O, T$\}. Here $R$ corresponds to a scenario determined as representative of all the usage scenarios belonging to the concept
(discussed below). $S$ corresponds to the rest of the usage scenarios in the bucket that we wrap under a `See Also'
sub-list. $O$ is the overall star rating for the bucket
(discussed below). $T$ is the title of the concept. We describe the fields
below.      

\begin{enumerate}
\item {Representative Scenario (R).} We pick the most recent scenario as the most representative. 
This decision is based on previous findings of 178 developers from SO and GitHub~\cite{Uddin-SurveyOpinion-TSE2019} 
that developers give a code example more value that is more recent (based on the post creation time). One major reason for this is that APIs undergo rapid version changes. As such 
the code examples that are provided even six months ago can be obsolete/less useful for the newer version of the API. The ranking of API usage scenarios using other 
metrics (e.g., scores) or a combination of scores and recency could be some other alternatives, which we leave for future work. 
\item {Concept Title (T).} We assign to the concept the title of the most
representative usage scenario.
\item {See Also (S).} We put the rest of the usage scenarios in the concept in a `See Also' bucket. 
\item {Concept Star Rating (O).} We compute the overall five star rating of the
concept by taking into account all the positive and negative reactions towards the usage
scenarios. For every concept of an API, we collect all the positive and negative opinions 
from all of its scenarios, and compute the star rating using \eq\ref{eq:star-rating}.
%
\end{enumerate}

\subsection*{Javadoc Adaptation. Type-Based Documentation} \label{sec:type-summary} 
We generate the type-based documentation of an API by grouping the scenarios of the API based on the API type as follows.
\begin{inparaenum}
\item We identify all the types in a code example associated with the API
\item We create a bucket for each type and put each code example in the bucket
where the type was found. We present this bucket as a sorted list by time (most
recent as the topmost).
\end{inparaenum} Our type-based documentation approach is similar to Baker~\cite{Subramanian-LiveAPIDocumentation-ICSE2014}, who proposed to annotate the Javadoc of an API 
class using all the code examples in SO where the API class was found. For example, based on \fig\ref{fig:HowAPIsDiscussed}, the code example will be put under two classes of the Google GSON API (Gson and TypeToken), 
one class under java.util API (List) and one class under java.lang API (Type).  The class `Data' is ignored, because it is a locally declared class in the code example. 

%

\subsection{Opiner Architecture to Support the API Documentation Algorithms}\label{subsec:generate-deploy-opiner}
We implemented the API usage scenario documentation algorithms by extending Opiner~\cite{Uddin-OpinerReviewAlgo-ASE2017}. We focused 
on following two requirements during the extension: 
\begin{enumerate}
\item Scalability. The Opiner's system should be able to process millions of
posts from developer forums.
\item Efficiency. The processing should be completed in short time, ideally in
hours.
\end{enumerate}
\begin{figure}
\vspace{-2mm}
\centering
\includegraphics[scale=0.73]{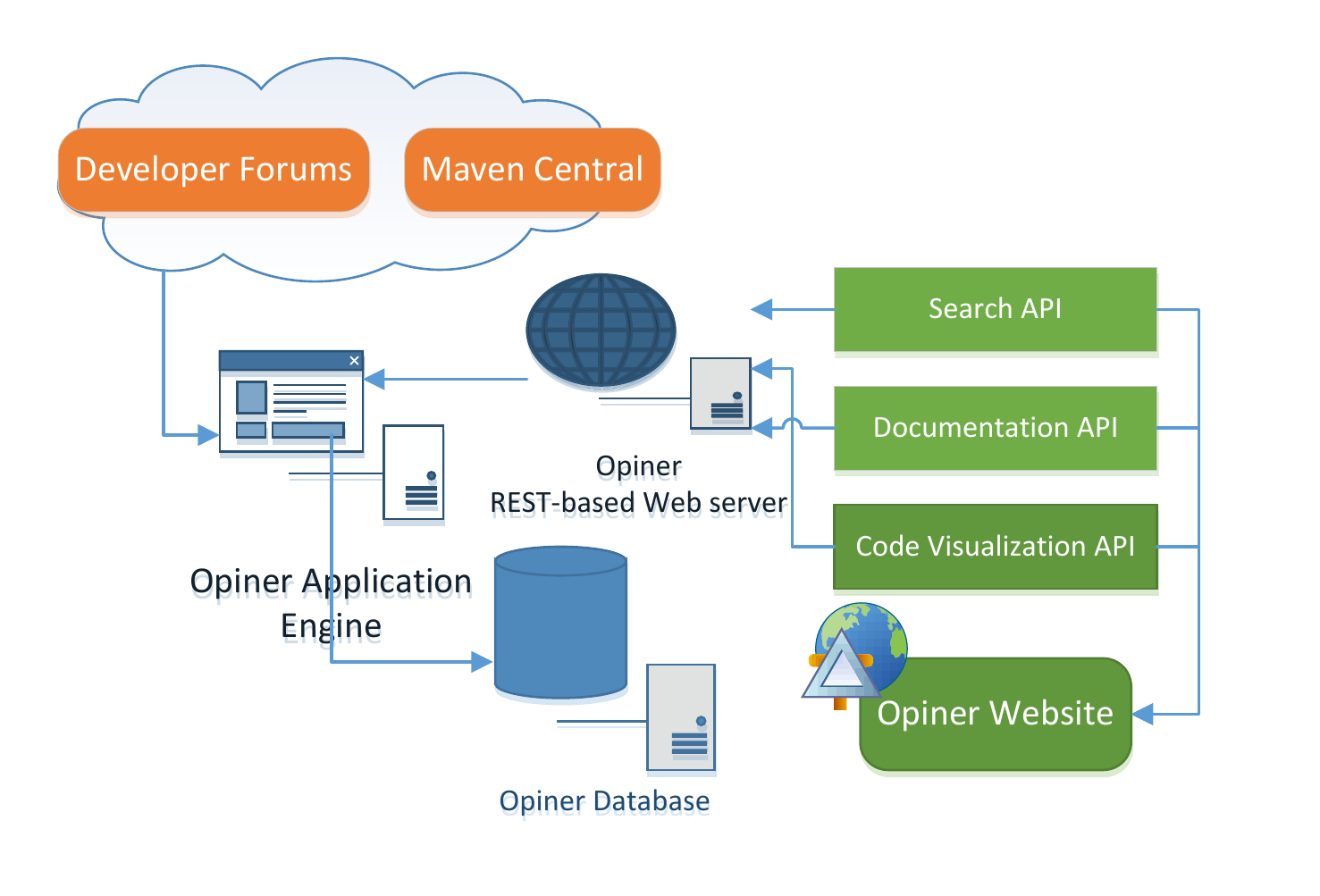}
\vspace{-5mm}
\caption{The system architecture in Opiner to support the proposed API usage scenario documentation algorithms}

\label{fig:opiner-arch}
\vspace{-5mm}
\end{figure} The
Opiner's system architecture is able to
process millions of posts. The proposed documentation algorithms are supported in Opiner by four major components
(see \fig\ref{fig:opiner-arch}):
\begin{enumerate}
  \item {Database:} A hybrid data storage component to manipulate data
  in diverse formats.
  \item {Application Engine:} Handles the loading and processing of the data
  from forum posts and hosts all the algorithms to mine and
  summarize usage scenarios about APIs.
  \item {REST Web Server:} The middleware between the Application Engine and
  the Website that supports two types of REST APIs:
  \begin{inparaenum}
  \item Search. to find an API with auto-completion support.
  \item Documentation. provides the documentation and visualization produced by the three algorithms.
  \end{inparaenum}
  \item {Website:} The web-based user interface to host the
  search and documentation results for API usage scenarios.
\end{enumerate}
The decoupling of the `Application Engine' from other components allows it to
run independently of the website. Thus the `Application Engine' can run offline,
even when the Website is being used. The `Application Engine' is designed to
load and preprocess each SO thread in parallel. The documentation
algorithms are applied on the preprocessed post contents. Our profiling of the
`Application Engine' shows that the preprocessing of the contents takes almost
80\% of all the time. This is because this step handles the creation of the meta
information of posts based on preprocessing, such as the parsing of code
snippets, the linking of code snippets to the API mentions, for examples. The
documentation component takes less than 10\% of the time. Therefore, once API usage scenarios 
are mined and properly preprocessed, the generation of the documentation using the proposed algorithms 
should take minimal time in Opiner.

\subsection{API Usage Scenario Documentation in Opiner Web Site}\label{sec:documentation-opiner}
We applied the algorithms on a dataset consists of 3,048 threads from SO that contain a total of 22.7K posts (question+answer+comment). 
The Opiner online web-site 
currently indexes the mined and documented usage scenarios of the APIs found in this dataset. 
The threads in the dataset are tagged as `Java+JSON', i.e., the posts discussed about JSON-based tasks using Java APIs. This 
dataset was previously used in \cite{Uddin-OpinerReviewAlgo-ASE2017} to summarize reviews about APIs and was found to offer 
a rich set of competing APIs with diverse API usage discussions. As such, we expect to see 
code examples discussing about Java APIs for JSON parsing 
in the posts. This dataset contains numerous usage scenarios from multiple competing APIs to support JSON-based manipulation in Java (e.g., REST-based architectures, microservices, etc.). 
JSON-based techniques can be used to support diverse development
scenarios, such as, serialization of disk-based and networked files, lightweight communication
between server and clients and among interconnected software
modules, messaging format over
HTTP as a replacement of XML, in encryption techniques, and on-the-fly conversion
of Language-based objects to JSON formats, etc.

We parsed the dataset to collect all the code examples. There are 8,596
\it{valid} code snippets (e.g., Java code) and 4,826 invalid code snippets (e.g., XML block). 
In \tbl\ref{tbl:datasets-overview} we show descriptive statistics of the dataset. A total of around 15K API names are mentioned in the textual contents of the posts. 
On average, each valid snippet contained at least 7.9 lines.   
The last column ``Users''  in \tbl\ref{tbl:datasets-overview} shows the total number of distinct users that posted at least one
answer/comment/question. On average, around
four users participated in one thread. More than one user participated in
2,940 threads (96.4\%). A maximum of 56 distinct users participated in one
thread~\cite{website:stackoverflow-338586}. 

\begin{table}[t]
\caption{Descriptive statistics of the dataset used to produce API documentation in Opiner website}
\centering
\begin{tabular}{rr|rr|rr|r|r}\toprule
\textbf{Threads} & \textbf{Posts} & \textbf{Sentences} & \textbf{Words}  &
\textbf{Snippet} & \textbf{Lines}   & \textbf{APIs Mentioned in Texts} & \textbf{Users}\\
\midrule
3,048 & 22.7K & 87K & 1.08M & 8,596  &  68.2K  & 15,605 & 7.5K\\ \midrule
\textbf{Average} & 7.5 & 28.6 & 353.3  & 2.8 & 7.9 & 5.1 & 3.9\\
\bottomrule
\end{tabular}
\label{tbl:datasets-overview}
\end{table}

\begin{table}[t]
  \centering
  \caption{Distribution of Code Snippets By APIs}
    \begin{tabular}{rrrr|rrrr}\toprule
    \multicolumn{4}{c}{\textbf{Overall}} & \multicolumn{4}{c}{\textbf{Top 5}} \\
    \midrule
    {\textbf{API}} & {\textbf{Snippet}} 
    & {\textbf{Avg}} & {\textbf{STD}} 
    & {\textbf{Snippet}} & {\textbf{Avg}} 
    & {\textbf{Max}} & {\textbf{Min}} \\
    \midrule
    175   & 8,596  & 49.1  & 502.7 & 5,196  & 1,039.2 & 1,951  & 88    \\
    \bottomrule
    \end{tabular}%
  \label{tbl:api-resolution-overview}%
\end{table}%
\begin{table}[t]
  \centering
  \caption{Distribution of Reviews in Scenarios with at least one reaction}
    \begin{tabular}{lrr|rr|rr}\toprule
    \textbf{Scenarios} & \multicolumn{2}{c}{\textbf{Comments}} & \multicolumn{2}{c}{\textbf{Positive}} & \multicolumn{2}{c}{\textbf{Negative}} \\
    \midrule
    \bf{w/Reviews} & {\textbf{Total}} & {\textbf{Avg}} &{\textbf{Total}} & {\textbf{Avg}} & \multicolumn{1}{l}{\textbf{Total}} & \multicolumn{1}{l}{\textbf{Avg}} \\
    \midrule
    1,154  & 7,538  & 6.5   & 2,487  & 2.2   & 1,216  & 1.1   \\
    \bottomrule
    \end{tabular}%
  \label{tab:api-comment-overview}%
\end{table}%

The 8,596 code examples are associated with 175
distinct APIs using the code example to API mention linking algorithm from \sec\ref{sec:mining-component}. 
\tbl\ref{tbl:api-resolution-overview} presents the distribution of the APIs by the code examples. 
The majority (60\%) of the code examples were associated to five
APIs for parsing JSON-based files and texts in Java: java.util, org.json,
Google Gson, Jackson, and java.io. Some API types are  
more widely used in the code examples than others. For example, the \tt{Gson} class from Google Gson API was found in 679 code examples out of the 
1,053 code examples linked to the API (i.e., 64.5\%). Similarly, the \tt{JSONObject} class from the org.json API was found in 
1,324 of 1,498 code examples linked to the API (i.e., 88.3\%). Most of those code examples also contained other types of the APIs. 
Therefore, if we follow the documentation approach of Baker~\cite{Subramanian-LiveAPIDocumentation-ICSE2014}, we would 
have at least 1,324 code examples linked to the Javadoc of \tt{JSONObject} for the API org.json. This is based on the parsing of 
our 3,048 SO threads. Among the API usage scenarios in our study dataset, we found 1,154 scenarios contained at least one review using 
our proposed algorithm to associate reviews to an API usage scenario. 
In \tbl\ref{tab:api-comment-overview}, we show the distributions of comments and reviews in the 1,154 scenarios. There are a total of 7,538 comments found in the corresponding 
posts of those scenarios, out of which 2,487 are sentences with positive polarity and 1,216 are sentences with negative polarity.     

%% file: evalAgainstTypeSummary.tex
\section{Usefulness of Proposed Crowd-Sourced API Documentation Algorithms}\label{sec:evalProposedAgainstType}
In this section, we present the results of a user study that we conducted to compare the usefulness of our two proposed documentation algorithms (Statistical and Concept-based) 
over type-based documentation of API usage scenarios. 
\subsection{Motivation}
Previous research find that a Javadoc-style hierarchical API documentation format is less useful to developers than a more practical task-centric documentation~\cite{Shull-InvestigatingReadingTechniquesForOOFramework-TSE2000,Uddin-HowAPIDocumentationFails-IEEESW2015}. 
As previously reported, 
effective developers investigate source code by looking for structural 
cues~\cite{Robillard-HowEffectiveDevelopersInvestigateSourceCode-TSE2004} and such exploration could involve the 
usage of multiple types of a single API~\cite{Uddin-TemporalApiUsage-ICSE2012}. Thus, a traditional type-based 
documentation may not handle the more useful task-based documentation format as hypothesized by Carroll et al.~\cite{Carroll-MinimalManual-JournalHCI1987a} as 
later confirmed by Shull et al.~\cite{Shull-InvestigatingReadingTechniquesForOOFramework-TSE2000}. 
We thus need to understand whether and how our proposed documentation algorithms 
could offer added benefits over a traditional type-based documentation format across diverse development scenarios. 
Such assessment can be formed based on inputs from software developers on the documentation produced in Opiner using the three algorithms.

\subsection{Approach}
We investigate how software developers rate the three documentation types (our two proposed + type-based) 
in Opiner along four development scenarios. Our \it{goal} was to judge the \it{usefulness} 
of a given documentation as shown in Opiner (see \sec\ref{sec:documentation-opiner}). The
\it{objects} were the three types documentation produced for a given API using the algorithms and the
\it{subjects} were the participants who rated each documentation. The \it{contexts}
are four development scenarios. The scenarios are previously used in Uddin and Khomh~\cite{Uddin-OpinerReviewAlgo-ASE2017}. 

\subsubsection{Participants.} We recruit 29 software developers. Among the 29 participants, 18 were professional 
developers. The rest of the participants (11) were recruited from four universities: 
two in Canada (University of Saskatchewan and Polyt\'{e}chnique Montreal) and
two in Bangladesh (Bangladesh University of Engineering \& Technology and
Khulna University). 
The 18 professional developers were recruited through the online professional social network site,
Freelancer.com. Sites like Amazon Mechanical turks and Freelancer.com have
been gaining popularity to conduct studies in empirical software engineering
research due to the availability of efficient,
knowledgeable and experienced software engineers. In our study,
we only recruited a freelancer if he had professional software development experience in Java. Among the 11 participants recruited from the universities, eight reported their
profession as students, two as graduate researchers. Among the 18 freelancers, one was a
business data analyst, four were team leads,
and the rest were software developers.
Among the 29 participants 88.2\% were actively involved in software development
(94.4\% among the freelancers and 81.3\% among the university participants).
Each participant had a background in computer science and software engineering.

The number of years of experience of the participants in software development
ranged between less than one year to more than 10 years: three (all of them being students) 
with less than one year of experience,
nine between one and two, 12 between
three and six, four between seven and 10 and the rest
(nine) had more than 10 years of experience.
Among the four participants
that were not actively involved in daily development activities, one was a
business analyst (a freelancer) and three were students
(university participants). The business data
analyst had between three and six years of development experience in Java. The
diversity in the participant occupation offered us insights into whether and how
Opiner was useful to all participants in general.

\subsubsection{Tasks.} We asked the participants to compare the documentation produced by the three
algorithms under four different development scenarios
(selection, documentation, presentation, and API authoring). 
Each task was described using a hypothetical development scenario where the
participant was asked to judge the summaries through the lens of a software
engineering professional. Persona based usability studies have been proven
effective both in Academia and Industry~\cite{Mulder-PersonaStudy-2006}. We
describe the tasks below.

\begin{enumerate}[leftmargin=10pt]
  \item \ul{Selection.} \it{Can the usage documentation
help you to select this API?} The persona was a `Software Architect' who was
tasked with making decision on the selection of an API given the usage documentation
produced for the API in Opiner. The participants were asked to consider the
following decision criteria in their answers: the documentation
\begin{inparaenum}[(C1)]
\item contained all the \it{right} information,
\item was \it{relevant} for selection, and
\item was \it{usable}.
\end{inparaenum}
\item \ul{Documentation.} \it{Is the produced documentation complete and readable?} 
The persona was a `Technical API Documentation  Writer' who was tasked
with the writing of the documentation of an API by
taking into accounts the usage summaries of the API in Opiner. The decision
criteria on whether and how the different summaries in Opiner could be useful
for such a task were: \begin{inparaenum}[(C1)]
\item the \it{completeness} of the information, and
\item the \it{readability} of the summaries.
\end{inparaenum}
\item \ul{Presentation.}  \it{Can the documentation easily help you to justify your selection
of the API?} The persona was a development team lead who was tasked with the
creation of a presentation by using the summaries in Opiner to justify the
selection of an API. The decision
criteria were:
\begin{inparaenum}[(C1)]
\item the \it{conciseness} of the information and
\item \it{recency} of the provided scenarios.
\end{inparaenum}.
\item\ul{Authoring.} \it{Can the documentation easily help you to decide whether to improve
an API feature?} the persona was an API author who was tasked with the creation of
a new API by learning the strengths and weaknesses of the competing APIs using Opiner. The decision criteria were:
\begin{inparaenum}[(C1)]
\item the reactions towards code examples and
\item  the presence of \it {diverse
scenarios}.
\end{inparaenum}
\end{enumerate}
We assessed the ratings of the three tasks (Selection, Documentation,
Presentation) using two metrics: useful (the documentation does not miss any info,  
misses some info but still useful), not useful (misses all the info so not useful at all). For the task (Authoring), we define usefulness as follows: 
useful (Fully helpful, helpful), not useful (Partially Unhelpful, Fully
unhelpful). 
For the authoring scenario, we further asked the participants whether he had
decided to author a new API as a competitor to the Jackson API whose documentation he had
analyzed in Opiner. The options were: Yes, No, Maybe. Jackson is the most popular Java API for JSON parsing. Each participant was 
asked to justify his rating for each scenario in a text box. 
The study was conducted in a Google form.

\subsection{Results}

\begin{table}[t]
  \centering
  \caption{Impact of the summaries (in percentages) based on the scenarios}
    \begin{tabular}{rlrrr}\toprule
    \multicolumn{1}{l}{\textbf{Scenario}} & \textbf{Rating} & \multicolumn{1}{l}{\textbf{Type-Based}} & \multicolumn{1}{l}{\textbf{Statistical}} 
    & \multicolumn{1}{l}{\textbf{Concept-Based}} \\ \midrule
    \multicolumn{1}{l}{\textbf{Selection}} & Useful & 93.1  & 93.1  & \textbf{100.0} \\
          & Not Useful & 6.9   & 6.9   & 0 \\
          \midrule
    \multicolumn{1}{l}{\textbf{Documentation}} & Useful & 93.1  & 79.3  & \textbf{100.0} \\
          & Not Useful & 6.9   & 20.7  & 0 \\
          \midrule
    \multicolumn{1}{l}{\textbf{Presentation}} & Useful & 93.1  & \textbf{96.6} & 93.1 \\
          & Not Useful & 6.9   & 3.4   & 6.9 \\
          \midrule
    \multicolumn{1}{l}{\textbf{Authoring}} & Useful & 72.4  & 69.0  & \textbf{89.7} \\
          & Not Useful & 10.3  & 10.3  & 3.4 \\
          & Neutral & 17.2  & 20.7  & 6.9 \\ 
          \bottomrule
    \end{tabular}%
   \label{tbl:scenario_overview_usefulness}
\end{table}%

In \tbl\ref{tbl:scenario_overview_usefulness}, we show the percentage of the
ratings for each usage documentation algorithm for the four development scenarios
(Selection, Documentation, Presentation, and Authoring). 

\addtocounter{o}{1}
\begin{tcolorbox}[flushleft upper,boxrule=1pt,arc=0pt,left=0pt,right=0pt,top=0pt,bottom=0pt,colback=white,after=\ignorespacesafterend\par\noindent]
\nd\it{\bf{Observation \arabic{o}.}} Our two proposed documentation algorithms (Concept and Statistical) were rated as more useful than the type-based documentation (Javadoc adaptation) across all four 
development scenarios.  
\end{tcolorbox} 

For the `Selection' scenario, the  incorporation of reactions as positive and negative
opinions in the usage documentation were considered as useful. According to one
participant: \emph{``Conceptual documentation is the most useful of all. Inexperienced
developers can select code snippet based on positive or negative reactions while the
experienced developers can compare the code and go for the best one.} However, just
the presence of the reactions were not considered as enough for selection of an
API when coding tasks were involved.  
\emph{``Statistical documentation just shows the negative and positive views, it is
not useful to view the working examples or code. Conceptual documentation groups
together the common API features, it helped me to find the common examples in
same place. Type based documentation is also ok but had to dig in to find the
description. All code examples provided the full code example, so it was good.''}

For the `Documentation' scenario, the participants
appreciated the innovative presentation formats of clustering usage scenarios by concepts. According to one
participant \emt{``For Documentation purpose , Conceptual documentation is more
readable than other two documentation.''}. Similar to the observations of Carroll et al.~\cite{Carroll-MinimalManual-JournalHCI1987a} for
the needs for documenting APIs based on tasks, the participants advocated the potential of concept-based summaries documentation,
such as \emt{Conceptual documentation is important to generate different kind of ideas
and thoughts to complete different kind of task such as serialization, deserialization, format specification, 
mapping with Jackson API. This documentation is most useful to me because of the availability of resource.}.
The participants considered type-based documentation to be useful for specific tasks that
may not involve multiple types of a given API at the same time.

For the `Presentation' scenario, the participants preferred the visualized charts from statistical documentation. Their suggested workflow was to 
start the presentation with the charts and then dive deeper into the presentation based on insights from concept-based documentation. According to one participant, the
combination of ratings and examples is the key: \emt{If I was the team lead,
statistical documentation would help to decide me to view the users positive and
negative reaction over the selection API, I would definitely take this into
consideration. Conceptual documentation would help me to create a presentation based
on the examples and the ratings.} The participants asked for an extension in statistical documentation to compare
APIs by the features offered, \emt{The statistical documentation helps to summarize the popularity of the API and other
co-existing APIs. But it would be more helpful if there was comparisons among
competitive APIs as well. The team lead should should this information one by one.}

For the `API Authoring' scenario, when asked about whether the documentation of the Jackson API in
Opiner showed enough weakness of the API that the developers would like to
author a competing API, 48.4\% responded with a `No', 35.5\% with a `Yes' and
16.1\% with Maybe.
The participants considered the concept-based documentation to be most useful during such decision making, followed by statistical documentation.  
According to one participant, \emt{In order to author a new API, I would like to understand how the market of developers
is reacting to the current API. More negative responses would indicate that there is
dissatisfaction among the developers and there is a need to create a new API for which I would look into the negative responses in the conceptual documentation.
Given that I can see that there is a positive trend for jackson Api from statistical documentation and there are no or very 
few negative responses from the users among the eight usage scenarios that I had selected.
I feel that I would not author a new API to compete Jackson, simply because it still works and the developers are content, 
which I could identify from the statistical documentation.} 
\addtocounter{o}{1}
\begin{tcolorbox}[flushleft upper,boxrule=1pt,arc=0pt,left=0pt,right=0pt,top=0pt,bottom=0pt,colback=white,after=\ignorespacesafterend\par\noindent]
\nd\it{\bf{Observation \arabic{o}.}} The concept-based documentation was considered as the most useful in three out of the four scenarios (Selection 100\%, Documentation 100\%, 
and Authoring 89.7\%), while the statistical documentation was considered as the most useful for the other scenario (Presentation 96.6\%). 
\end{tcolorbox}

%% file: evalCoding.tex
\section{Effectiveness of Opiner Documentation Over API Web Documentation Resources}\label{sec:evaluation}
In this section, we evaluate the effectiveness of the documentation in Opiner web site against 
traditional API web documentation resources, i.e., API official documentation and developer forum (SO).  
\subsection{Motivation}
The goal of the automatic documentation of APIs usage scenarios in
Opiner is to assist developers in finding the right solutions to their coding
tasks with relative ease than other resources. Therefore, we need to assess the effectiveness of Opiner API usage documentation in real-world coding tasks. 
Previous research reported that developers consider the combination of code examples and API reviews in the online developer forums 
as a form of API documentation~\cite{Uddin-SurveyOpinion-TSE2019} and preferred such documentation over API official documentation. However, the developers 
struggled to make a complete and concise insights due to the huge volume and scattered nature of the usage discussions in the online forums. Therefore, the usage scenario documentation in Opiner 
should alleviate the pains of developers to find a complete solution to a development task without going over multiple SO posts. 
 Previous research 
also showed that the API documentation is often incomplete, obsolete and incorrect~\cite{Uddin-HowAPIDocumentationFails-IEEESW2015} and that 
developers find it hard to learn and use an API by simply relying on API official documentation~\cite{Robillard-APIsHardtoLearn-IEEESoftware2009a}.  
Therefore, the usage scenario documentation in Opiner should be able to compensate for such shortcomings in the official documentation of an API during their coding tasks.

\subsection{Approach}
We recruited 31 developers and asked them to complete four coding tasks using the documentation produced in Opiner and using API 
web documentation resources (i.e., baselines). At the end of the coding tasks, we asked the participants to participate in a short survey to share their experience 
of using Opiner over the baseline resources.  The participants used the documentation produced in Opiner (see \sec\ref{sec:documentation-opiner}) for the coding tasks.


\subsubsection{Participants.} The coding tasks were completed by a total of 31 participants. 29 out of the 31 participants came from our previous 
study to evaluate the usefulness of our API documentation algorithms in \sec\ref{sec:evalProposedAgainstType}. 
The additional two participants in this study are recruited from universities. Both of them are 
graduate students with more than one year experience in software development. 
Among the 31 participants, 18 were professional 
developers. The rest of the participants (13) were recruited from four universities. 
Each freelancer was remunerated with \$20, which was a modest sum
given the volume of the work. Each participant had a background in computer science and software engineering.
The survey questions were answered by 29 out of the
31 participants.

\subsubsection{Coding Tasks.} The four tasks are described
in \tbl\ref{tbl:coding-tasks}.
Each task was required to be completed using a pre-selected API.
Thus for the four tasks, each participant needed to use four different APIs:
Jackson~\cite{website:jackson}, Gson~\cite{website:gson},
Xstream~\cite{website:xstream}, and Spring framework~\cite{website:spring}.
Jackson and Gson are two most popular Java APIs for
JSON parsing~\cite{website:stackoverflow-2378402}.
Spring is one of the most popular web framework in
Java~\cite{website:javaFrameworkPopularity} and XStream is well-regarded
for its efficient adherence to the
JAXB principles~\cite{website:stackoverflow-1558087}, i.e.,
XML to JSON conversion and vice versa. 

All of the four APIs can be found in the list of top 10 most discussed APIs in our evaluation corpus. The APIs are mature and are fairly large and thus can be hard
to learn. The largest API is the Spring framework 5.0.5 with a total of 3687
types (Class, Annotation, etc.), followed by Jackson 2.2 (467
types), XStream 1.4.10 (340 types), and Gson 2.8.4 (34 types). 
For Jackson API, we counted the code types that are shipped with the core
modules (jackson-core, databind, and annotations). The Jackson API has been
growing with the addition of more modules (e.g., jackson-datatype). The four APIs
have a total of 4528 types. In comparison, the entire Java SE 7 SDK has a total of
4024 code types, and the entire Java SE 8 SDK has a total of 4240 code
types.\footnote{We used the online official Javadocs of the APIs to collect the
type information. Our online appendix contains the list.}

\subsubsection{Task Settings.} For each coding task, we prepared four
settings:
\begin{description}[leftmargin=2em]
\item[\bf{SO}]complete task using only SO,
\item[\bf{DO}]complete task using only API official documentation,
\item[\bf{OP}]complete the task only with the help of Opiner, and
\item[\bf{EV}]complete task using any online resources available (i.e., SO, DO, Opiner, and Search engines).
\end{description} The above four settings help us properly analyze the diverse coding behavior of software developers using 
the diverse online API documentation resources available to complete a coding task. Intuitively, the EV setting is the most natural, because 
it allows a developer to use search engines and/or any other resources. The SO, DO, and OP settings simply restrict the access of resources. 
This restriction then provides insights whether the participants are more/less useful while simply using a subset of all available resources. 
If, a user still shows performance similar/higher similar to the EV setting while only using Opine (i.e, the OP setting), that then can offer increased confidence on the 
the overall effectiveness of Opiner to support developers in their coding tasks.

We follow a between-subject
design~\cite{Woh00}, with four different groups (G1,
G2, G3, G4) of participants,
each participant completing four tasks in four different settings.
Each of three groups
(G1, G3, G4) had eight participants. At the end, eight participant from the three groups (G1, G3, G4) and seven from G2 completed the coding tasks. 
Each participant in a group was asked to complete the four tasks. Each
participant in a group completed the tasks in the order and settings shown in \tbl\ref{tbl:task-distribution}. 
To ensure that the participants used the correct development resource for a given API in a given development setting, we
added the links to those resources for the API in the task description. For example, for the task TJ and the setting SO, 
we provided the following link to query SO using Jackson as a tag:
\urls{https://stackoverflow.com/questions/tagged/jackson}. For the task TG and the setting DO, we provided the following link to the official documentation page of the Google GSON API:
\urls{https://github.com/google/gson/blob/master/UserGuide.md}. For the task TX
and the setting PO, we provided a link to the summaries of the API XStream in
the Opiner website
\urls{http://sentimin.soccerlab.polymtl.ca:38080/opinereval/code/get/xstream/}.
For the task TS with the setting EV, we provided three links, i.e., one from
Opiner (as above), one from SO (as above), an one from API formal
documentation (as above). For the `EV' setting, we added in the instructions that the participants are free to use any search engine.

To avoid potential bias in the coding
tasks, we enforced the homogeneity of the groups by
ensuring that:
\begin{inparaenum}
\item no group entirely contained participants that were only professional
developers or only students,
\item no group entirely contained participants from a specific geographic
location and--or academic institution,
\item each participant completed the tasks assigned to him independently and
without consulting with others
\item each group had same number of four coding tasks
\item each group had exposure to all four development settings as part of the
four development tasks.
\end{inparaenum} The use of balanced groups simplified
and enhanced the statistical analysis of the collected data.


\subsubsection{Task Selection Rationale.} The four tasks were picked randomly from our evaluation corpus of 22.7K SO posts. 
Each task was observed in SO posts more than once and was asked by more than one
developer. Each task was related to the manipulation of JSON inputs using Java
APIs for JSON parsing. The manipulation of JSON-based inputs is prevalent in
disk-based, networked as well as HTTP-based message, file, and object
processing.  Therefore, the Java APIs for JSON parsing offer many complex
features to support the diverse development needs.
The solution to each task spanned over two posts. The two posts are from two
different threads in SO. Thus the developers could search in Stack
Overflow to find the solutions. However, that would require those searching posts
from multiple threads in SO. All of those tasks are common using the
four APIs.
Each post related to the tasks was viewed and upvoted more than hundred times in
SO. To ensure that each development resource was treated with
\it{equal fairness} during the completion of the development tasks, we also made
sure that each task could be completed using any of the development resources,
i.e., the solution to each task could be found in any of the resources at a
given time, without the need to rely on the other resources.


\begin{table}[tbp]
\caption{Overview of coding tasks}
\begin{tabular}{llp{12.5cm}}\toprule

\textbf{Task} & \textbf{API} & \textbf{Description} \\ \midrule TJ & Jackson &
Write a method that takes as input a Java Object and serializes it to Json,
using the Jackson annotation features that can handle custom names in Java
object during deserialization. \\
\midrule TG & GSON & Write a method that takes as input a JSON string and
converts it into a Java Object. The conversion should be flexible enough to
handle unknown types in the JSON fields using generics. \\
\midrule TX & Xstream & Write a method that takes as input a XML string and
converts it into a JSON object. The solution should support aliasing of the
different fields in the XML string to ensure readability \\
\midrule TS & Spring & Write a method that takes as input a JSON response and
converts it into a Java object. The response should adhere to strict JSON
character encoding (e.g., UTF-8). \\
\bottomrule
\end{tabular}
\label{tbl:coding-tasks}
\end{table}

%
%

\begin{table}[tbp] \centering
\caption{Distribution of coding tasks per group per setting. 
SO = SO, DO = Javadoc, OP = Opiner, EV = Everything. TJ, TG, TX, TS
= Task Using Jackson, GSON, XStream, Spring, resp.}
\begin{tabular}{lrrrr}\toprule

 \bf{$\downarrow$ Group $|$ Setting $\rightarrow$ } & \textbf{SO} & \textbf{DO} & \textbf{OP} &
 \textbf{EV}
 \\
 \midrule
\textbf{G1} & TJ & TG & TX & TS \\
\textbf{G2} & TS & TJ & TG & TX \\
\textbf{G3} & TX & TS & TJ & TG \\
\textbf{G4} & TG & TX & TS & TJ \\
\bottomrule
\end{tabular}
\label{tbl:task-distribution}
\end{table}

\subsubsection{Coding Guide.} A seven-page coding guide was produced to
explain the study requirements (e.g., the guide for Group G1: \urls{https://goo.gl/ccJMeY}).
Before each participant was invited to complete the study, he had to read the entire coding guide. Each participant was encouraged to ask questions to clarify the study details before and during the study. To respond to the questions the participants communicated with the first author over Skype.
Each participant was already familiar with formal and informal documentation
resources. To ensure a fair comparison of the different resources used to complete the
tasks, each participant was given a brief demo of Opiner
before the beginning of the study. This was done by giving them an
access to the Opiner web site.

\subsubsection{Data Collection Process.} The study was performed in a
Google Form, where participation was by invitation only.
Four versions of the form were generated, each corresponding to one group.
Each group was given access to one version of the
form representing the group. An offline copy of each form is provided in our
online appendix~\cite{website:opinerusagescenariotse-online-appendix}.
The developers were asked to write the solution to each coding task in
a text box reserved for the task. The developers were encouraged to use
any IDE to code the solution to the tasks. 

Before starting each task, a participant
was asked to mark down the beginning time.
After completing a solution the participant was again asked to
mark the time of completion of the task. 
The participant was encouraged to not take any break during the completion of
the task (after he marked the starting time of the task). 
To avoid fatigue, each participant was encouraged to
take a short break between two tasks.
Besides completing each coding task, each participant was also asked to assess
the complexity and effort required for each task,
using the NASA Task Load Index (TLX)~\cite{Hart88},
which assesses the subjective workload of subjects. After completing each task,
we asked each subject to provide their self-reported effort on the completed task
through the official NASA TLX log engine at \url{nasatlx.com}.
Each subject was given a login ID, an experiment ID and task IDs,
which they used to log their effort estimation for each task, under the different settings.

\subsubsection{Variables to Assess The Coding Tasks.} The main independent variable we consider is the development
resource that participants use to find solutions for their coding tasks. 
Dependent variables are the values of the following metrics: correctness of the code solutions, time, and effort spent to code the solutions (discussed below). 

%


\begin{inparaenum}
\item \ul{Correctness.} To check the correctness of the solution for a coding task,
we used the following process:
\begin{inparaenum}
\item We identified the correct API elements (types, methods) used for the coding task.
\item We matched how many of those API elements were found in the coding solution and in what order.
\item We quantified the correctness of the coding solution using the following
equation:
\begin{equation}\label{eq:correctness}
\tr{Correctness} = \frac{|\tr{API Elements Found}|}{|\tr{API Elements
Expected}|}
\end{equation}
\end{inparaenum}
An API element can be either an API type (class, annotation) or an API method.
Intuitively, a complete solution should have all the required API elements expected for the solution.
We discarded the following types of solutions: \begin{inparaenum}
\item \it{Duplicates.} Cases where 
the solution of one task was copied into the solution of another task. 
We identified this by seeing the same solution copy pasted for the two tasks. 
Whenever this happened, we discarded the second solution. 
\item \it{Links.} Cases where developers only provided links to an online 
resource without providing a solution for the task. We discarded such solutions.
\item \it{Wrong API.} Cases where developers provided the solution using an API that
was not given to them. 
\end{inparaenum}

\item\ul{Time.} We computed the time taken to develop solutions for each task, by
taking the difference between the start and the end time reported for the task by the participant. 
Because the time spent was self-reported, it was prone to errors (some participants failed to record their time correctly). 
To remove erroneous entries, we discarded the following type of
reported time: \begin{inparaenum}
\item reported times that were less than two minutes. It takes time to read the
description of a task and to write it down, and it is simply not possible to do all such activities within a minute.
\item reported times that were more than 90 minutes for a given task.
For example, we discarded one time that was reported as 1,410 minutes, i.e.,
almost 24 hours. Clearly, a participant cannot be awake for 24 hours to
complete one coding task. This happened in only a few cases.
\end{inparaenum}

\item \ul{Effort.} We used the TLX metrics values as reported by the
participants. We analyzed the following five dimensions in the TLX metrics
for each task under each setting:
\begin{inparaenum}[\bfseries(a)]
\item \it{Frustration Level.} How annoyed versus complacent the developer
felt during the coding of the task?
\item \it{Mental Demand.} How much mental and perceptual activity was
required?
\item \it{Temporal Demand.} How much time pressure did the participant feel
during the coding of the solution?
\item \it{Physical Demand.} How much physical activity was required.
\item \it{Overall Performance.} How satisfied was the participant with his
performance?
\end{inparaenum} Each dimension was reported in a 100-points range with 5-point
steps. A TLX `effort' score is automatically computed as a task load index by
combining all the ratings provided by a participant. 
Because the provided TLX scores were based on the judgment of the participants, they are prone to subjective bias. 
Detecting outliers and removing those as noise from such ordinal data is a standard statistical process~\cite{Tukey-ExploratoryDataAnalysis-Pearson1977}. 
By following Tukey, we only considered values between the following two ranges as valid:
\begin{inparaenum}
\item Lower limit: First quartile - 1.5 * IQR
\item Upper limit: Third quartile + 1.5 * IQR
\end{inparaenum}.
Here IQR stands for `Inter quartile range', which is calculated as: $\tr{IQR} =
Q3 - Q1$. Q1 and Q3 stand for the first and third quartile, respectively.
\end{inparaenum}

\subsubsection{Follow Up Survey.} The survey was conducted in a Google doc form. We asked five questions, the first three questions were related to 
API official documentation and the last two were related to SO:
\begin{enumerate}
  \item How do the Opiner documentation offer improvements over formal documentation?
  \item How can Opiner 
  complement formal API documentation? 
  \item How would you
  envision for Opiner to complete the formal API documentation?
  \item How do Opiner summaries offer improvements over the
  SO contents for API usage?
  \item How would you envision Opiner to complement
  SO in your daily development activities?
\end{enumerate}
The first and the fourth question had a five-point Likert scale (Strongly Agree, Agree, Neutral, Disagree, and Strongly
Disagree). The other questions were answered in text boxes. 

\subsection{Results from Coding Tasks}
A total of 135 coding solutions
were provided by the participants. We discarded
14 as invalid solutions (i.e., link/wrong API). Out of the 135
reported time, we discarded 23 as being spurious.
24 participants completed the TLX metrics
(providing 96 entries in total), with each setting having six participants.
\begin{table}[t]
  \centering
   \caption{Summary statistics of the correctness (scale 0 - 1), time (minutes) and effort (Scale 0-100) spent in the
   coding tasks. Baselines: SO = SO, DO = Official Documentation, EV = Everything including search engine}
    \begin{tabular}{lr|rr|rr|rr}\toprule
          & \multicolumn{1}{l}{\textbf{Opiner (OP)}} & \multicolumn{1}{l}{\textbf{SO}} 
          & $\Delta_{SO, OP}$ & \multicolumn{1}{l}{\textbf{DO}} 
          & $\Delta_{DO, OP}$ & \multicolumn{1}{l}{\textbf{EV}} & $\Delta_{EV, OP}$ \\
          \midrule
    \textbf{Correctness} & \textbf{0.62} & 0.46  & -35\%$\downarrow$  & 0.5   & -24\%$\downarrow$  & 0.55  & -13\%$\downarrow$ \\
    \textbf{Time} & \textbf{18.6} & 22.3  & 17\%$\uparrow$ & 23.7  & 22\%$\uparrow$ & 19.4  & 4\%$\uparrow$ \\
    \textbf{Effort} & \textbf{45.8} & 55.8  & 18\%$\uparrow$ & 63.9  & 28\%$\uparrow$ & 54.8  & 16\%$\uparrow$ \\
    \bottomrule
    \end{tabular}%
  \label{tbl:coding-task-summ-stat}%
\end{table}%
\begin{figure*}
\centering
\includegraphics[scale=0.45]{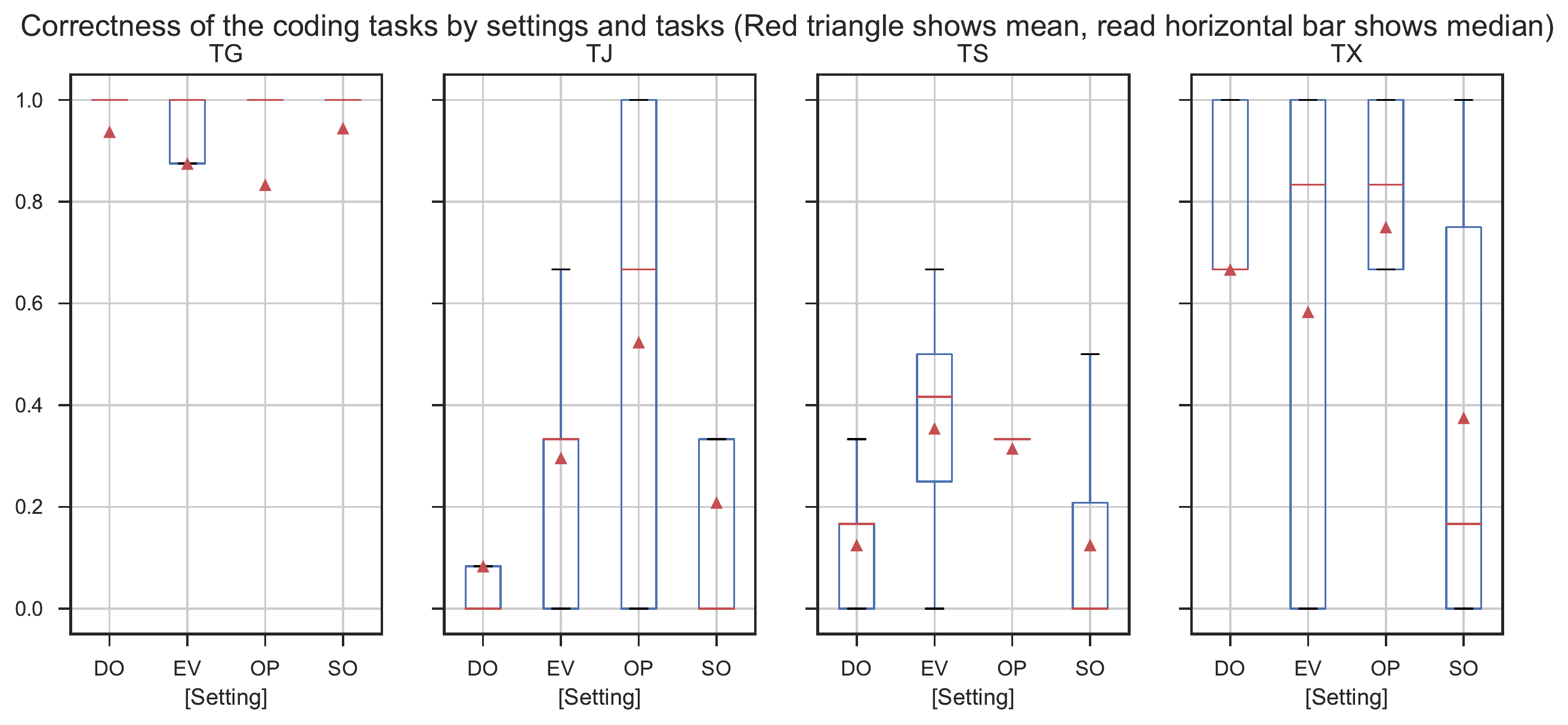}
\includegraphics[scale=0.45]{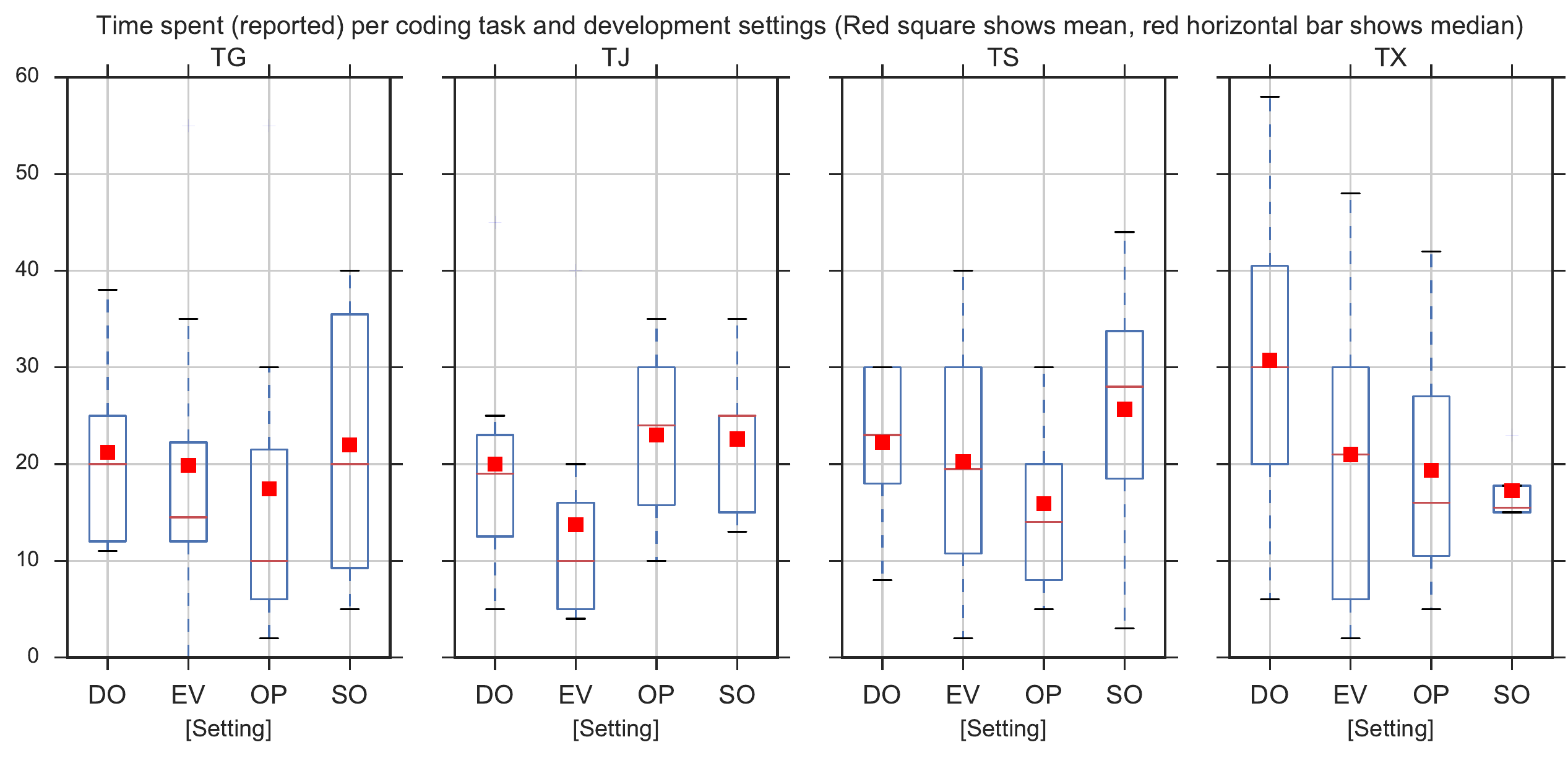}
\includegraphics[scale=0.45]{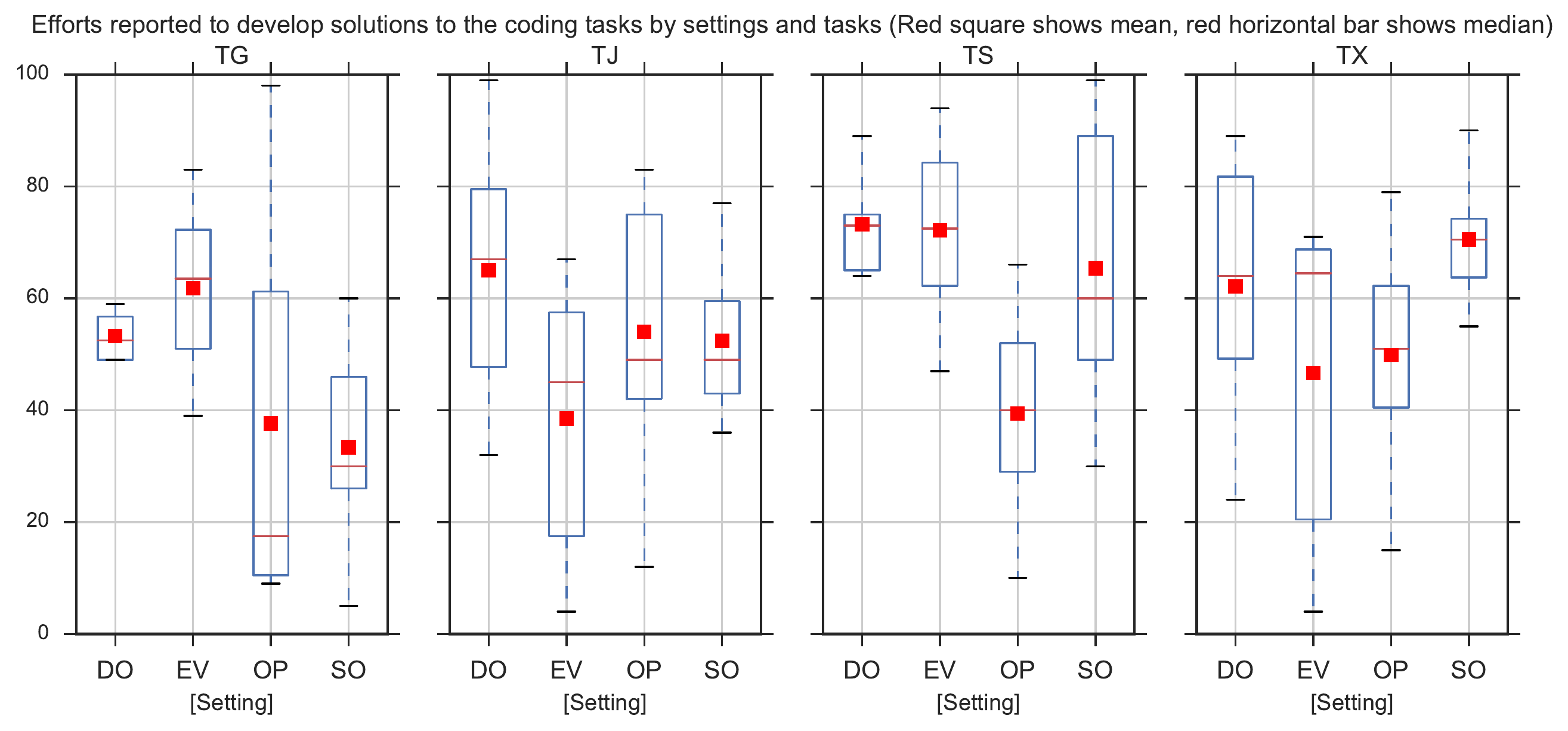}
\caption{Correctness of the completed solutions and the time and effort spent by tasks. Red horizontal bar shows the median}
\label{fig:boxblot-correctness}
\end{figure*}

\tbl\ref{tbl:coding-task-summ-stat} compares Opiner against the three baselines (API official documentation, SO, and Everything including search engines) 
based on the three metrics: 
the average correctness of the provided solutions, and the average time and effort spent to complete the solutions. 
The columns with $\Delta$ computes the percent difference between Opiner and each baseline for each metric. For example, $\Delta_{SO, OP}$ for the metric `Effort' is computed as:
\begin{equation}
\Delta_{Effort_{SO}, Effort_{OP}} = \frac{{Effort_{SO} - Effort_{OP}}}{Effort_{SO}} 
\end{equation} Recall that the effort calculation uses the NASA TLX software effort index. 

\addtocounter{o}{1}
\begin{tcolorbox}[flushleft upper,boxrule=1pt,arc=0pt,left=0pt,right=0pt,top=0pt,bottom=0pt,colback=white,after=\ignorespacesafterend\par\noindent]
\nd\it{\bf{Observation \arabic{o}.}} {Opiner outperforms each baseline for each metric. 
The participants coded with the maximum correctness (0.62), with least time (18.6 minutes) and effort (45.8) per coding solution 
using Opiner.} 
\end{tcolorbox} 
%
%
Among the three baselines, the `EV' setting was the most useful for the developers. The EV setting contained everything (SO, API official documentation, 
Opiner, and search engine), i.e., the developers were able to consult all the documentation resources available in the Web. {Opiner still outperformed 
the `EV' setting; developers coded with less correctness (-13\%), and spent more time (4\%) and effort (16\%) while using `EV' setting than Opiner.} 
\addtocounter{o}{1}
\begin{tcolorbox}[flushleft upper,boxrule=1pt,arc=0pt,left=0pt,right=0pt,top=0pt,bottom=0pt,colback=white,after=\ignorespacesafterend\par\noindent]
\nd\it{\bf{Observation \arabic{o}.}} {The correctness of the provided solutions were
the lowest when participants used only SO.} {For code correctness, the difference is the maximum between Opiner and SO (35\% less correct code while using SO)}. 
\end{tcolorbox} 
The difficulty of developers to produce a correct solution just by relying on SO confirms previous findings that 
developers 
are facing difficulty while attempting to find the right solution from the millions of forum posts~\cite{Uddin-OpinerReviewAlgo-ASE2017,Uddin-SurveyOpinion-TSE2019}. 
\addtocounter{o}{1}
\begin{tcolorbox}[flushleft upper,boxrule=1pt,arc=0pt,left=0pt,right=0pt,top=0pt,bottom=0pt,colback=white,after=\ignorespacesafterend\par\noindent]
\nd\it{\bf{Observation \arabic{o}.}} Participants on average spent the highest
amount of time and effort per coding solution
when using only the formal documentation. {For both effort and time spent, the differences are the 
maximum between Opiner and API formal API documentation (22\% more time and 28\% more effort while using API official documentation).} 
\end{tcolorbox} 
The difficulty of developers while just relying on formal API documentation confirms previous findings~\cite{Robillard-FieldStudyAPILearningObstacles-SpringerEmpirical2011a,Shull-InvestigatingReadingTechniquesForOOFramework-TSE2000} 
that developers can be less productive while using API official documentation, because it is hard to develop solutions quickly due to the various shortcomings in API 
official documentation (e.g., incompleteness, ambiguity, presentation)~\cite{Uddin-HowAPIDocumentationFails-IEEESW2015}. 

In \fig\ref{fig:boxblot-correctness}, we show the correctness of the coding tasks, effort and time spent across the four
settings using four boxplots, one for each coding task. The red square in each
box shows the mean ($\mu$), and the red horizontal line shows the median ($M$) of accuracy. For example, for the task TJ, the mean in
accuracy using Opiner was 0.52 while the median was 0.67. The participants showed more accuracy while using 
Opiner for two tasks (TJ and TX) than while using the other resources. For the task TG, they showed lower accuracy ($\mu = 0.83$)
while using Opiner than while using othe resources. In fact, they achieved almost perfect accuracy for the task TG,
while using both SO and API official documentation. For the task TS, while the participants showed the
highest accuracy using every resources (i.e., EV setting with an average $\mu = 0.39$ and median $M = 0.45$), Opiner was the second best with a
slightly lower mean accuracy of 0.38 and a median of 0.4.

{The participants spent less
time while using Opiner, than the other resources for the two tasks (TG,
TS). For the task TX, participants spent on average 19.4 minutes using Opiner was the 
second best behind (17.3 minutes) only SO (the lowest completion
time).} Out of all tasks, the participants spent the longest time (on average)
while completing the task TJ across all the settings. Even though the participants spent the most amount of time (on average) 
to complete the task TJ, they achieved the best accuracy for the task while using Opiner. 

The participants spent the least amount of effort while using Opiner for one task (TS). 
For the task (TJ), developers spent more effort when using the official documentation than Opiner.
\rev{Task TJ required using the Jackson API. The official documentation of
Jackson is considered to be complex by developers in SO~\cite{website:stackoverflow-q2378402}. Intuitively, the clustering of
similar usage scenarios in Opiner could be useful to find relevant code examples
easily. Indeed, the participants in the survey reported that they found the
concept-based documentation of Jackson in Opiner useful to complete the tasks.} For the task
TS involving the Spring Framework, Opiner considerably outperformed the other settings for this 
task (less than 40\% for Opiner to more than 70\% for formal documentation).
\begin{table}[t]
  \centering
   \caption{Results of the Wilcoxon Mann-Whitney U test and Cliff's delta effect size for the
   pair-wise comparisons of the accuracy, time, and effort spent per the four settings (OP = Opiner only, DO = Official documentation only, SO = Stack Overflow only, 
   EV = Everything including search engine) for the four coding tasks}
   \begin{tabular}{llrr}\toprule
    \multicolumn{1}{l}{\textbf{Metric}} & \textbf{Comparison} & \multicolumn{1}{l}
    {\textbf{$p-value$}} & \textbf{Cliff's $\delta$} \\
    \midrule
    \multicolumn{1}{l}{Correctness} & \bf{OP vs DO} & 0.08 & 0.206 S \\
          & \bf{OP vs SO} & 0.04* & 0.265 S \\
          & \bf{OP vs EV} & 0.27 & 0.092 N \\
          \cmidrule{2-4}
          & DO vs SO & 0.31 & 0.075 N \\
          & DO vs EV & 0.21 & 0.118 N \\
          & SO vs EV & 0.12 & 0.174 S \\
          \midrule
    \multicolumn{1}{l}{Time} & \bf{OP vs DO} & 0.12 & 0.206 S \\
          & \bf{OP vs SO} & 0.08 & 0.244 S \\
          & \bf{OP vs EV} & 0.47 & 0.015 N \\
          \cmidrule{2-4}
          & DO vs SO & 0.43 & 0.030 N \\
          & DO vs EV & 0.10 & 0.221 S \\
          & SO vs EV & 0.08 & 0.246 S \\
          \midrule
    \multicolumn{1}{l}{Effort} & \bf{OP vs DO} & 0.03* & 0.330 S \\
          & \bf{OP vs SO} & 0.15 & 0.177 S \\
          & \bf{OP vs EV} & 0.14 & 0.181 S \\
          \cmidrule{2-4}
          & DO vs SO & 0.18 & 0.158 S \\
          & DO vs EV & 0.28 & 0.101 N \\
          & SO vs EV & 0.32 & 0.080 N \\
          \bottomrule
    \end{tabular}%
  \label{tbl:rq1-nonparametrictests}%
\end{table}%

\rev{In \tbl\ref{tbl:rq1-nonparametrictests}, we compare the three variables (accuracy per coding solution, time and effort spent per solution) by the four settings (OP = Opiner only, DO = Official documentation only, SO = Stack Overflow only, 
   EV = Everything including search engine). We use  Wilcoxon Mann-Whitney U test to determine whether the difference between any two settings for a given variable 
   is statistically significant (i.e, $\alpha \le 0.05$). Mann-Whitney U test is non-parametric, which is used in software engineering literature to compare data that 
   is not normally distributed~\cite{Hu-MobileAppConsistency-EMSE2018}. We estimate the effect size between the distributions. We use cliff's delta which is suitable for non-parametric 
   distribution (unlike Cohen delta which is suitable for normal distribution). Cliff's delta produces a value between -1 and 1. Following Romano et al.~\cite{Romano-TtestCohenD-SAIR2006}, we use the following threshold 
   to interpret the cliff's delta effect size:}
   \begin{equation}
  \tr{Effect Size} =\left\{
  \begin{array}{@{}ll@{}}
    {negligible~(N)}, & \tr{if}\ |\delta| \leq 0.147 \\
    {small~(S)}, & \tr{if}\ 0.147 < |\delta| \leq 0.33 \\
    {medium~(M)}, & \tr{if}\ 0.33 < |\delta| \leq 0.474 \\
    {large~(L)}, & \tr{if}\ 0.474 < |\delta| \leq 1 
  \end{array}\right.
\end{equation} \rev{The effect size categories (negligible, small, medium, 
or large) quantify the differences between the distributions. For the coding accuracy variable, the difference between 
OP and SO is statistically significant, i.e, the participants were significantly more 
effective to produce correct coding solutions while using Opiner only than Stack Overflow only. 
For the effort spent, the difference between OP and DO is statistically significant, i.e., 
the participants spent significantly less time to complete their coding tasks while using Opiner than 
while using the API official documentation. The effect sizes are small or negligible, given that 
the size our dataset is not large enough.} 
 
\addtocounter{o}{1}
\begin{tcolorbox}[flushleft upper,boxrule=1pt,arc=0pt,left=0pt,right=0pt,top=0pt,bottom=0pt,colback=white,after=\ignorespacesafterend\par\noindent]
\nd\it{\bf{Observation \arabic{o}.}} The participants reported the most satisfaction about their coding solution while using Opiner for three out the four tasks (TJ, TS, TX). 
For the other task (TG), Opiner was slightly behind official documentation. With TG, developers found all solutions in one single Javadoc page, which contributed to its 
higher satisfaction score.  
\end{tcolorbox} 

\nd\bf{Opiner vs EV.} \rev{Despite allowing the participants to use any resources available to complete a
coding task in the EV setting, the participants spent least time and effort but
with the most accuracy while coding using Opiner only. We did not have any
follow up questions to the participants to understand this phenomenon. To understand the
reasons behind this, we simulated the four coding tasks under the EV setting.
We sought to accept or refute the following hypothesis:}
\rev{\begin{enumerate}[label=\bf{H\arabic{*}}]
  \item The Search engine did not have the right
  results to complete the coding task.
  \item The participants may not have used Search engine at all, despite being asked
  to do otherwise (i.e., when a solution was indeed present in the Search Engine
  result).
\end{enumerate}}
\rev{For each task, we searched the Google search engine on May 6, 2018. The search
query was a string that combined the API name with the task description 
as shown in \tbl\ref{tbl:coding-tasks}. Therefore, for the task involving the
Jackson API (i.e., TJ), the query was ``Jackson + $<$Description$>$''. We
manually analyzed each of the results returned in the first page of Google. For each
result, we searched for code examples in the web page linked to the result. If
one or more code examples are found in the linked web page, we assessed the
accuracy of the code examples by computing the coverage of API elements expected
in the correct solution against the found code examples, i.e., using
\eq\ref{eq:correctness}. If a solution was not found in the search result, we
accept the first hypothesis (H1). If a solution was found, we further compare
the found solution against the solutions coded by the participants. If the
comparison shows no or low coverage we accept the second hypothesis (H2). For a
search result pointing to a question in SO, we took the solution
found in the accepted answer of the question (following Subramanian et
al.~\cite{Subramanian-LiveAPIDocumentation-ICSE2014}). If no accepted answer is
found, we discarded that result.} 
 
\begin{table}[t]
  \centering
  \caption{The perceived correctness of the solutions using results from
  the first page of Google}
    \begin{tabular}{lrrrrrr}\toprule
    \textbf{Task} & \multicolumn{1}{l}{\textbf{Links}} & \multicolumn{1}{l}{\textbf{Mean}} & \multicolumn{1}{l}{\textbf{SD}} & \multicolumn{1}{l}{\textbf{Median}} & \multicolumn{1}{l}{\textbf{Max}} & \multicolumn{1}{l}{\textbf{Min}} \\
    \midrule
    TJ (Jackson) & 5     & 0.42  & 0.28  & 0.5   & 0.67  & 0 \\
    TG (Gson) & 10    & 0.25  & 0.4   & 0     & 1     & 0 \\
    TX (Xstream) & 9     & 0.19  & 0.28  & 0     & 0.67  & 0 \\
    TS (Spring) & 10    & 0.1   & 0.2   & 0     & 0.5   & 0 \\
    \midrule
    Overall Average & 8.5   & 0.24  & 0.29  & 0.125 & 0.71  & 0 \\
    \bottomrule
    \end{tabular}%
  \label{tab:google-simulation}%
\end{table}%
\rev{In \tbl\ref{tab:google-simulation}, we show the \it{perceived} correctness of
the solutions for the coding tasks based on the Google search result. The second
column (Links) shows the number of search links provided for each task as a
search query. The last five columns provide descriptive statistics (e.g.,
average, median) of the correctness of the solutions to coding tasks using the
code examples found in the links. On average, we got 8.5 hits (i.e, links) per search query (i.e., API
name + Task description). The number of hits was the lowest for the task involving Jackson API. 
However, the average correctness was the highest for this API. For Spring, the
number of hits is highest (jointly with Gson), but the average correctness was
the lowest. The major reason of the lower correctness is that most links provide
code examples involving other APIs. Some links did not have any code example at
all. For example, one hit for the Spring API only consists of all the different
ways the Spring framework can be configured.} 

\rev{Overall, the average correctness of the coding tasks is higher while using
Opiner than while using Google search result only (see
\tbl\ref{tbl:coding-task-summ-stat} for statistics about Opiner). We observed
the following reasons for this phenomenon:}
\rev{\begin{enumerate}
\item\bf{Discoverability.} Google is not a code search engine. Therefore, many
links did not have any code examples. There is no explicit filter in Google
search engine to explicitly specify that we need the results only for an API and
not for something else (for example, Spring can be many things besides being a
Java framework, such as a season, a store, a perfume, etc.). Both of these two
mitigating factors were not helpful to search for coding solutions using Google.
\item\bf{Scatteredness.} The solutions to each of our four coding tasks were
scattered in more than one SO post. Using Google search result, we
were able to find half of the solutions in many cases. For example, the maximum
correctness using the Spring API was only 50\% due to that reason. 
\end{enumerate} Despite this, we note that when Google offered good results, it was indeed very
good. For example, the maximum correctness using the Gson API was 100\%, i.e.,
we were able create a complete solution simply by relying on Google search
result for the Gson API. Therefore, it would be interesting to see how Google
search could be combined with Opiner search to improve the overall developers'
experience.}

\subsection{Results from the Follow Up Survey}
\begin{figure}[t]
\vspace{-2.cm}
   	\includegraphics[scale=.7]{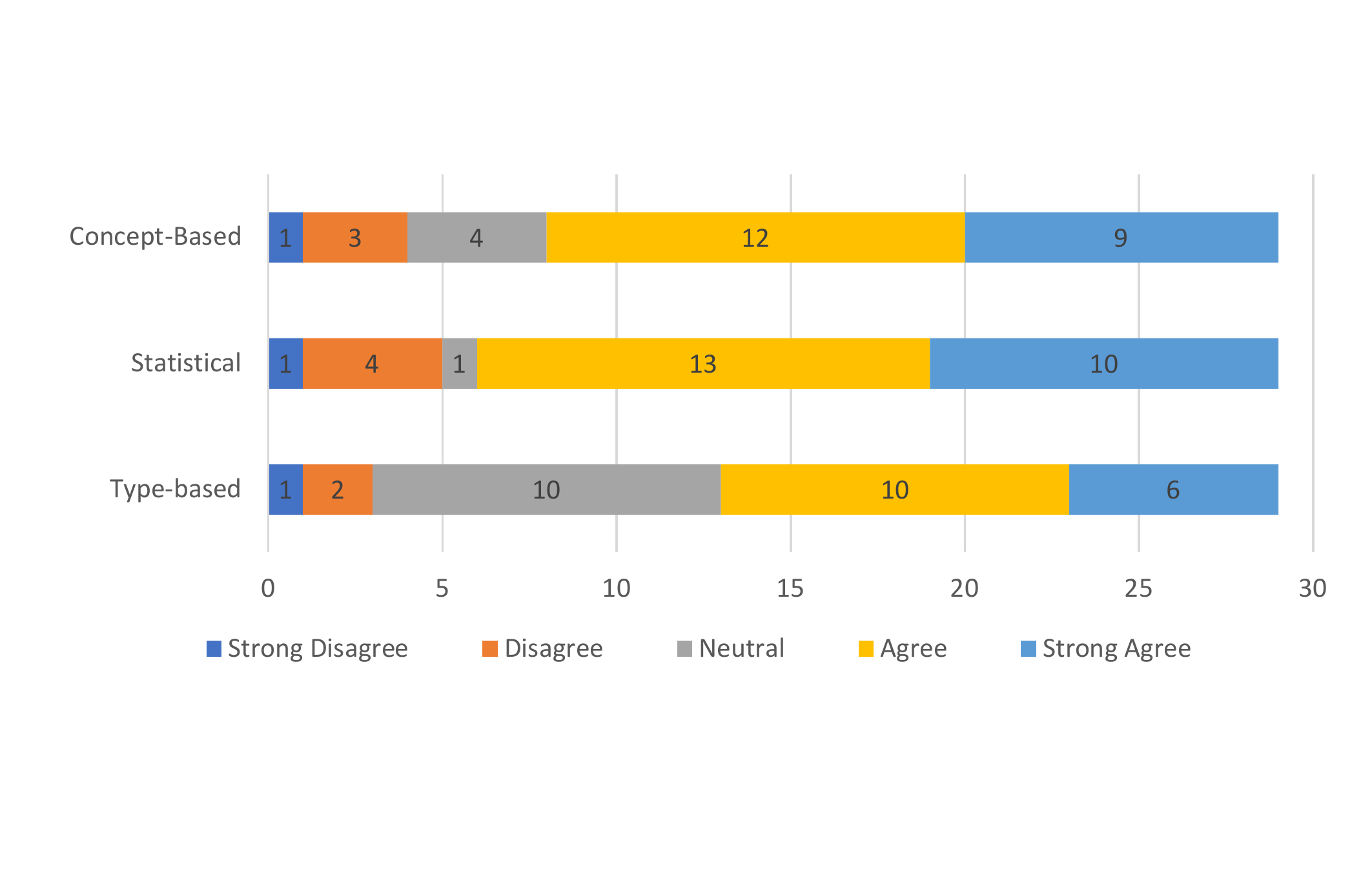}
   	\vspace{-2.cm}
   	 \caption{Developers' response to whether Opiner's usage documentation offer improvements over API official documentation}
   	 \label{fig:usefulness-of-summaries-over-javadoc}
\end{figure}

In \fig\ref{fig:usefulness-of-summaries-over-javadoc}, we show the responses of
the participants to the question whether Opiner offered improvements whether in their four development tasks over formal documentation and if so which documentation.
\addtocounter{o}{1}
\begin{tcolorbox}[flushleft upper,boxrule=1pt,arc=0pt,left=0pt,right=0pt,top=0pt,bottom=0pt,colback=white,after=\ignorespacesafterend\par\noindent]
\nd\it{\bf{Observation \arabic{o}.}} More than 85\% of the respondents agreed (= agree + strongly agree) that 
the new documentation algorithms in Opiner offered improvement over API official documentation in their four coding tasks.
\end{tcolorbox}
For the statistical and concept-based documentation, the participants appreciated the following improvements
in Opiner over API official documentation: \begin{inparaenum}
\item Uptodateness of the scenarios, and
\item Presence of sentiments to validate the effectiveness of usage scenarios.
\end{inparaenum} According to one participant, \emt{Statistical documentation would help me to find
the users decisions about the API. I would chose conceptual documentation 
rather than formal documentation because these
provides the usage example with positive and negative reaction.}
\addtocounter{o}{1}
\begin{tcolorbox}[flushleft upper,boxrule=1pt,arc=0pt,left=0pt,right=0pt,top=0pt,bottom=0pt,colback=white,after=\ignorespacesafterend\par\noindent]
\nd\it{\bf{Observation \arabic{o}.}} \rev{The respondents considered that
statistical and concept-based documentation complement the API official
documentation by offering innovative ways of documenting up to date API usage
scenarios with reviews from online developer forums.}
\end{tcolorbox}
The preference of the developers of our two proposed documentation algorithms over type-based documentation confirms previous 
seminal research by Carroll et al.~\cite{Carroll-MinimalManual-JournalHCI1987a} and 
Shull et al.~\cite{Shull-InvestigatingReadingTechniquesForOOFramework-TSE2000} that technical tasks are better supported by `minimal manual', i.e., 
where the produced documentation can help them easily and quickly to complete a given development task in hand with task-centric documentation focus.   
%


\addtocounter{o}{1}
\begin{tcolorbox}[flushleft upper,boxrule=1pt,arc=0pt,left=0pt,right=0pt,top=0pt,bottom=0pt,colback=white,after=\ignorespacesafterend\par\noindent]
\nd\it{\bf{Observation \arabic{o}.}} {Around 80\% of the participants agreed 
that Opiner documentation can complement API official documentation and as such the API documentation in Opiner (statistical and concept-based) should be integrated into 
the formal API documentation to produce a 
better official documentation.}
\end{tcolorbox}
According to one participant, \emt{Formal documentation will almost always be the
starting point for new APIs. But as the APIs start to grow, Opiner will serve as the most useful tool to 
find correct solutions to the problems in less time, become familiar with the trends and compare different alternatives.} 
In particular, the participants appreciated the combination of usage examples with discussions based on reactions in the concept-based documentation, 
\emt{One thing that official documentation miss on are the problems and discussion forum. Opiner could complement the formal documentation, 
if the conceptual documentation are incorporated within the API documentations.}

\begin{figure}[t]
   	\includegraphics[scale=.65]{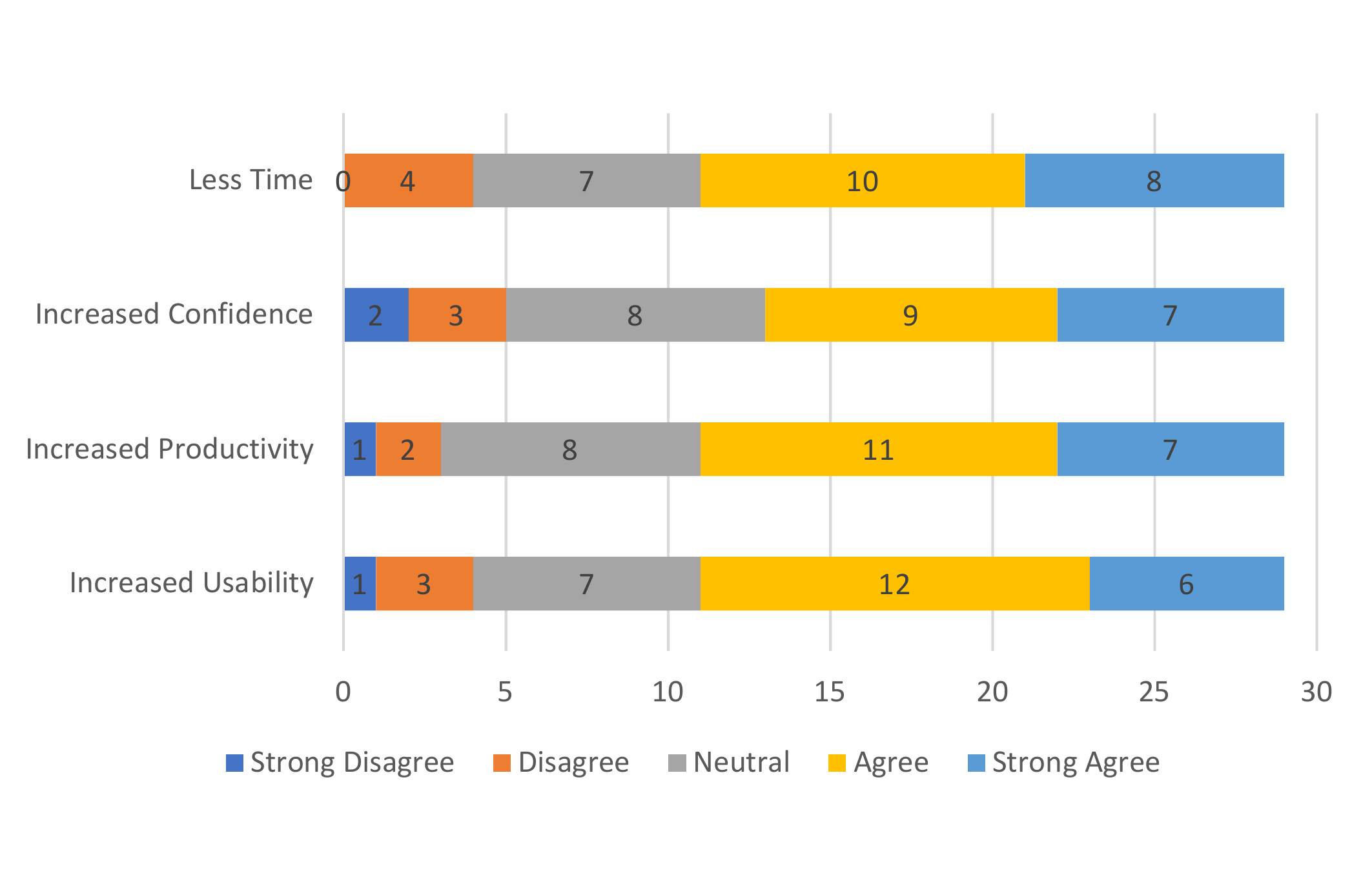}
   	\vspace{-1.cm}
   	 \caption{Developers' response to whether Opiner's usage documentation offer improvements over API informal documentation}
   	\vspace{-2mm}
   	 \label{fig:usefulness-of-summaries-over-forum}
\end{figure}



In \fig\ref{fig:usefulness-of-summaries-over-forum}, we show how developers
responded to our questions on whether in their coding tasks Opiner offered
benefits over the informal documentation. 
\addtocounter{o}{1}
\begin{tcolorbox}[flushleft upper,boxrule=1pt,arc=0pt,left=0pt,right=0pt,top=0pt,bottom=0pt,colback=white,after=\ignorespacesafterend\par\noindent]
\nd\it{\bf{Observation \arabic{o}.}} More than 80\% of the participants agreed that
Opiner documentation offered increased usability, productivity, confidence and saved time
over the informal documentation, such as SO. 
\end{tcolorbox}
According to one developer \emph{``there are some sections like conceptual
summary and code example would be useful for developer on daily basis''}. According to 
another participant \emph{``It is quicker to find solution in Opiner since the subject
is well covered and useful information is collected.''} The participants mentioned that Opiner 
offers more features than SO to learn an API: \emph{``Opiner has more feature set to help the developers
find out the better API for a task. The evaluation and nice presentation of positive and negative sentiments help them decide an API.''}
The participants wished for an AI-powered search feature in Opiner: \emph{``The most challenging thing I faced in tech based domains is the collection of my vocabulary. 
To search or ask something we need to first know the proper question. If we do not know with what terms I would ask something, then no only a tool but not even a human can answer me. ''}

\addtocounter{o}{1}
\begin{tcolorbox}[flushleft upper,boxrule=1pt,arc=0pt,left=0pt,right=0pt,top=0pt,bottom=0pt,colback=white,after=\ignorespacesafterend\par\noindent]
\nd\it{\bf{Observation \arabic{o}.}} With regards to complementarity of Opiner to SO, 
the participants considered learning an API using Opiner could be quicker than while using SO, 
because Opiner synthesizes the information from SO by APIs using both sentiment and source code analyses. 
\end{tcolorbox}

%% file: discussions.tex
\section{Implications}\label{subsec:implications}
The findings from our study can guide the following major stakeholders in software engineering: \begin{inparaenum}
\item \it{API Authors} to guide them during their creation of new APIs,  
\item \it{API Documentation Writers}  by offering them automated tool supports to produce API documentation,
\item \it{Software Developers} by offering them new and better API documentation format than traditional Javadocs, and 
\item \it{Software Engineering Researchers} to study new and innovative ways of documenting API usage scenarios from online technical Q\&A sites. 
\end{inparaenum} We discuss the implications below. 

\bf{\ul{API Authors.}} The number of open source repositories in GitHub was 10 million in 2013, 67 million in 2018, 100 million in 2019. This growth has been quite exponential. 
Within just last one year, 10 million new users joined GitHub contributing to 44 million repositories across every continent on earth~\cite{website:github-octoverse}. 
For an API author, this is a unique time to create and promote new APIs. However, the creation of an API can take significant time, resource and effort. 
The API documentation in Opiner can help them to make a decision on whether and how to develop a new competitor to an API. We observed this in our user study, when 
we asked the developers whether they would author a new competitor to the Java Jackson API. 
Out of the respondents 48.4\% responded with a `No' and 35.5\% responded with a `Yes'. 
The Statistical documentation page offers visualized insights into the various usage statistics of an API. 
Those who responded with a `No' referred to the more positive reviews around the usage scenarios of the API in Opiner: 
\emt{Given that I can see that there is a positive trend for Jackson API from statistical documentation and there are no or very 
few negative responses from the users among the eight usage scenarios that I had selected.
I feel that I would not author a new API to compete Jackson.} The concept-based documentation can show how conceptually relevant usage scenarios (i.e., concepts) are reviewed. For example, if a concept is negatively reviewed, 
the competitor API can focus more on improving the features related to the concept.  Those who responded with a `Yes' resonated: \emt{The negative sentiments in statistical 
summary indicate that many think it might not be best tool for JSON manipulation or there are other alternatives. 
It also indicate that an improved API can be derive for the same purpose. The conceptual summary help to identify the weaknesses.}   

\bf{\ul{API Documentation Writers.}} Without the presence of good
documentation and tutorials, it is difficult to learn any
API~\cite{Robillard-FieldStudyAPILearningObstacles-SpringerEmpirical2011a,Robillard-APIsHardtoLearn-IEEESoftware2009a}.
Unfortunately, official API documentation are often found to be incomplete,
ambiguous and even incorrect~\cite{Uddin-HowAPIDocumentationFails-IEEESW2015}. The problem with official documentation lies mainly 
on the lack of adequate resources to create and maintain the documentation~\cite{Robillard-FieldStudyAPILearningObstacles-SpringerEmpirical2011a}. This problem 
can be exacerbated when a developer moves to a new project landscape leaving behind the documentation he maintained in their previous projects, but now 
he needs to learn the new project by looking at the current project documentation~\cite{Dagenais-DeveloperLearningResources-PhDThesis2012,Dagenais-MovingToNewSoftwareLandscape-ICSE2010}.
Our proposed new API documentation algorithms in Opiner 
can assist API documentation writer to create new API documentation and to maintain existing documentation. 
This is because the documentation can be automatically created by analyzing how developers 
are discussing the different API usage scenarios in online technical Q\&A sites.  
In our survey, more than 80\% of the respondents agreed that the new documentation algorithms in Opiner offer improvement over official API documentation. 
The documentation writers can first generate the documentation using the algorithms and then decide which usage scenario to 
include in the official documentation based on the opinion analysis as follows: \emt{Conceptual documentation
presented the usage with rating, I would chose the answer with highest rating
cause it shows that this particular answer is correct on the given scenario, it
also has example plus more related answer.} The automated nature of our documentation algorithms can help documentation writers to keep the official documentation 
up to date.  In our survey, around 80\% of the participants agreed 
that Opiner documentation can complement API official documentation. The participants wished the 
the API documentation in Opiner to be integrated into 
the formal API documentation to produce a 
better official documentation.

\bf{\ul{Software Developers.}} The presence of good documentation is paramount for a software developer to be able to use an API~\cite{Robillard-APIsHardtoLearn-IEEESoftware2009a}.
In fact, developers often decide to stop using an API with low quality documentation~\cite{Uddin-HowAPIDocumentationFails-IEEESW2015}, or 
pick a competing API with a better documentation~\cite{Uddin-OpinionValue-TSE2019}. The proposed two documentation algorithms in Opiner can be used to 
create new API documentation and to complement existing API official documentation. This can help developers to select and use an API. While working in a team, 
the visualized presentation in Statistical documentation can be used to justify
the selection and usage of an API: \emt{The summary gives the positive and
negative sentiments regarding this API which is helpful for justifying the long
term goals.}. The clustering of similar API usage scenarios in concept-based documentation can help developers to dive deeper into conceptually relevant 
problems in an API: \emt{Generally the need to use a new API is engineered because of some issues with
existing one or some domain problem. Conceptual summary is best feature as it summarizes different relevant
problems and solutions. "See Also" is great feature. The sentiments on each
summary is great. Jumping
to the details is easy.}  

The developers in our studies considered the combinations of code examples and reactions (i.e., positive and negative opinions) was useful in Opiner and wished 
such format could be integrated into official documentation. This finding confirms our previous surveys of 178 developers, who reported to consider the 
combination of code example and API reviews as a form of API documentation~\cite{Uddin-SurveyOpinion-TSE2019}. Current research has focused on 
detecting API misuse patterns in SO code examples~\cite{Zhang-APIMisuseSOStudy-ICSE2018}. The detection of API misuse to inform developers of problematic code examples 
can be challenging due to the domain specific nature of the patterns (e.g., C++ patterns could be different from Java patterns). Our findings show that 
a combination of code example and reviews can offer a viable alternative to API misuse pattern detection, until we can reliably identify all different API misuse patterns in developer forums. 
\begin{figure*}[t]
    \centering
    \subfloat[Statistical Documentation]{
	\resizebox{2.5in}{!}{
    \begin{tikzpicture}

    \pie[explode=0.1, text=pin, number in legend, sum = auto, color={black!0, black!10, black!20, black!30, black!40}]
        { 17.2/\Huge{Every Day (17.2\%)},
          31/\Huge{Every Week (31\%)},
          31/\Huge{Every Month (31\%)},
          13.8/\Huge{Once a Year (13.8\%)},
          6.9/\Huge{Never (6.9\%)}
          }
    \end{tikzpicture}}%
    }
    \qquad
    \subfloat[Concept-Based Documentation]{
	\resizebox{2.5in}{!}{
    \begin{tikzpicture}

    \pie[explode=0.1, text=pin, number in legend, sum = auto, color={black!0, black!10, black!20, black!30, black!40}]
        { 34.5/\Huge{Every Day (34.5\%)},
          20.7/\Huge{Every Week (20.7\%)},
          31/\Huge{Every Month (31\%)},
          3.4/\Huge{Once a Year (3.4\%)},
          10.3/\Huge{Never (10.3\%)}
          }
    \end{tikzpicture}}%
    }
    \caption{Preferences of developers to use the Statistical and Concept-Based documentation in Opiner to stay aware}%
    \vspace{-0.35cm}
    \label{fig:prefStayAwareStatistical}%
\end{figure*}
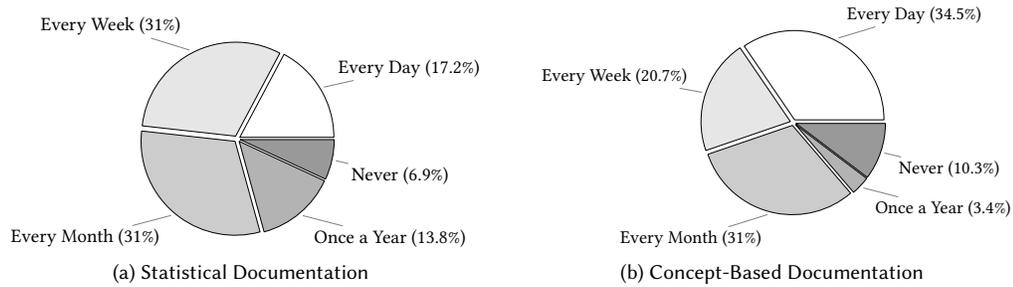

Finally, with its growing popularity, SO now has become quite huge. As such, developers 
are now finding it challenging to get quick but concise insights of of an API usage that can be scattered 
across millions of SO posts~\cite{Uddin-SurveyOpinion-TSE2019}. For the developers, the synthesized API documentation in Opiner can be useful. 
Indeed, more than 80\% of the developers in our survey agreed that
Opiner documentation offered increased usability, productivity, confidence and saved time
over the informal documentation, such as SO. In fact, when we asked the 29 survey participants whether they would like to stay aware of APIs by using our produced 
API documentation, more than 86\% mentioned that they would like to use Concept-based documentation at least once a month (34\% everyday) and more than 79\% mentioned that 
they would like to use Statistical documentation at least once a month (see \fig\ref{fig:prefStayAwareStatistical}).

\bf{\ul{Software Engineering Researchers.}} While software documentation is an established and active research for many decades, 
the automatic generation of API documentation from online Q\&A sites is a relatively new field. Existing such research has so far focused on 
augmenting API official documentation, such as Javadoc with code examples and interesting insights from SO~\cite{Subramanian-LiveAPIDocumentation-ICSE2014,Treude-APIInsight-ICSE2016}. 
Our study shows that we can offer innovative documentation options besides such javadoc adaptation. In our user study, our two proposed 
API documentation algorithms (Statistical and Concept-Based) were consistently preferred over Type-based documentation in all four development scenarios. This 
finding is consistent with previous seminal research of Carroll et al.~\cite{Carroll-MinimalManual-JournalHCI1987a} and Shull et al.~\cite{Shull-InvestigatingReadingTechniquesForOOFramework-TSE2000} who promoted task-centric 
documentation instead of traditional hierarchical documentation (e.g., javadoc). The participants in our study voiced the lack of usability 
with such a type-based format: \emt{In Type based summary I have to
search the documentation based upon the type, and then statistics and then the
solution. I would like it to be accessed easily.} Our proposed two new algorithms offered the study participants better usability and better access to 
right information in least time and least effort.  
Our findings thus 
can influence research that can combine human computer interaction with programming needs, such as incorporating useful visualized summaries into existing code completion 
tools~\cite{Zilberstein-Codota-ONWARD2016}, as well as crowd-sourced documentation techniques~\cite{Ponzanelli-PrompterRecommender-EMSE2014,Souza-CookbookAPI-BSSE2014}.
In summary, our study opens the door for new and innovative API documentation approaches, by promoting an effective departure from traditional documentation approaches.

%% file: threats.tex
	\section{Threats to Validity}\label{sec:threats}
We discuss the threats to validity of our studies following common guidelines
for empirical studies~\cite{Woh00}.

\bf{Construct Validity Threats} concern the relation between theory and
observations. In our study, they could be due to measurement errors. We assessed
the correctness of a provided solution to a coding task by comparing the
coverage of the API elements (types and methods) found in the solutions against those
as expected in a correct solution to the task. This approach is adopted from our
previous analysis of API usage concepts in API official
documentation~\cite{Uddin-TemporalApiUsage-ICSE2012}. This approach allows
us to measure the accuracy with a reproducible technique (i.e., using
\eq\ref{eq:correctness}). In particular, this approach is required to avoid the 
\it{Instrumentation} threat which arises when the measurement of the dependent
variable (e.g., correctness in our study) varies between the different groups of
participants.
We note, however, that a solution with all the expected API elements may still not be fully
correct, when they are not used properly (e.g., when the inputs are not set properly). 
We did not observe any such cases in the provided
solutions.

The accuracy of the algorithms to mine API usage scenarios can have an
impact on our produced documentation. In~\cite{Uddin-MiningAPIUsageScenarios-IST2020}, we 
report the performance of the algorithm. Each algorithm shows more than 80-90\% precision and recall. 
The algorithms outperformed eight state of the art baselines (see our published paper~\cite{Uddin-MiningAPIUsageScenarios-IST2020}).

We use clone detection during the identification of concepts in our proposed algorithm, Concept-based documentation. 
To detect clones, we applied a 60\% similarity. Svajlenko et al.~\cite{Svajlenko-EvaluatingCloneDetectionTools-ICSME2014} 
proposed this threshold to detect clones in big dataset, such as our dataset of more than 8K code examples. Given that the Concept-based 
documentation in Opiner is the most positively reviewed by the developers in our user study, the real-world application of the concepts 
shows promise. Further investigation of the clones may provide additional benefits, which we leave as future work. 

\rev{The atomic unit of analysis in our two proposed algorithms is an API usage scenario, which consists of a code example of an API, a textual description, and 
reviews towards the code example as found in the comments to the answer where the code example is found. Our 
user studies show that the study participants appreciated the reviews towards the code example and found those useful. 
However, research results show that not all comments can be insightful or informative~\cite{Zhang-ReadingAnswerSONotEnough-TSE2019}, 
such as comments those are upvoted by others and comments that are not disputed by others. 
Further improvements to our API documentation generation algorithms will focus on determining and showing informative comments. 
Nevertheless, the creation of our API usage scenarios benefits from reviews from comments because 
the reviews are useful to provide contextual information with regards to the usage of APIs. However, 
not all answers have many comments, i.e., in our dataset we could not add reviews to each API usage scenario. While we did not 
limit our study participants to only API usage scenarios with reviews, the experience of our study participants with more comments per 
usage scenarios could have been different and that could have influenced them to give even more preference to Opiner. Future extension of 
Opiner can be developed to monitor real-time any new comments posted against an answer and to collect comments from other answers 
that are similar to an answer.} 

\rev{Currently, we simply show all comments towards a code example by grouping the sentences from the comments into three polarity classes: 
sentences containing positive, negative, or neutral sentiments. However, some comments may already have been addressed into the 
provided example and thus may have become obsolete and/or misleading. Future extensions will compare the answer editing history to 
determine whether and how a comment is already addressed before we decide to include it in our API usage scenario.}    

\rev{Currently, an API usage scenario is assigned to one concept only in our proposed concept-based documentation algorithm (see \sec\ref{sec:documentation-component}). 
Intuitively, an API usage scenario can be linked to multiple concepts due to situational relevance between tasks that are addressed in two different concepts (i.e., 
soft assignment of API usage scenarios by allowing them to be included in multiple concepts). 
The concept-based documentation was the most favored by our study participants. An extension of the concept-based documentation with 
`soft' assignment of API usage scenarios) may make the concept-based documentation more useful to the developers. We leave it as our future work.}

\bf{Maturation Threats} concern about the changes in a participant during the
study due to the \it{passage of time}, such as development environment changes,
hunger, fatigue, etc. To
mitigate the effect of fatigue, we asked participants to take breaks between the
tasks. We also used TLX to measure the effort of participants. TLX is by its
nature subjective. However, it is a mature tool that has been used in multiple
empirical studies, e.g.,~\cite{Abbes11}. In addition, we discarded the
reported times by the participants in their coding tasks that are most likely
wrong, e.g., task completion time taking more than 24 hours (see
\sec\ref{sec:evaluation}).


\bf{Internal Validity Threats} compromise the confidence in attributing the
relationship between the dependent and independent variables. The
dependent variables in our coding tasks are the three metrics (correctness,
time and effort). The independent variables are the ratings of the
participants about Opiner and the baselines (e.g., the API documentation).
In particular \emph{Internal validity threats} relevant in this study are learning, selection,
and diffusion threats. 

\it{Learning/History threats} do not affect our study since we used
a between-subject design, which is a recommended approach in controlled study to
avoid such bias in the study~\cite{Woh00}. We followed standard
practices of between-subject studies in empirical software
engineering~\cite{Abbes11}: Each participant performed different tasks in different settings.  The tasks are
formulated in a way so that the learning from one task cannot be transferred 
to another task, even when the participants are using different APIs to complete the tasks (see Description column in \tbl\ref{tbl:coding-tasks}).
For example, the input to each task is different from each other. 
For the task involving Jackson (TJ) it is a Java object, for Gson it
is a JSON string, for XStream it is an XML string, and for Spring it is a JSON
response coming from the web server. Each task focused on the distinct features
offered by the APIs that distinguish the APIs from each other. Therefore, the
requirements with the inputs are also different: for Jackson the participant 
needs to use Jackson annotation feature to convert the Java object. For Gson, the participant needs
to handle unknown types in the input dynamically. For XStream the participant
needs to convert the input fields names to something custom using the aliasing
feature specifically provided by the XStream API. For Spring, the participant
needs to adhere to the specific encoding as specified in the JSON response.

\it{Diffusion threats} do not impact our study because our participants did not know each
 other, and therefore could not discuss about the experiments. 

\rev{\it{Selection threat} could arise if the participants were biased towards a specific
tool. The participants from our coding tasks were asked
to compare Opiner with the baselines, after completion of their coding tasks. 
While 87\% of the participants showed interest to use Opiner, almost half of
them responded with a `yes' and the other half with a `maybe'.  
The participants who responded with a `maybe' cited their lack of familiarity with Opiner, given the tool was 
introduced to them right before the coding tasks. In contrast, these developers were quite familiar with other 
resources (i.e., baselines), such as SO, official documentation and search engines.  
Several participants mentioned that
they would prefer SO over Opiner, because they are already familiar
with SO. However, we did not observe any bias towards SO or Opiner 
from the participants during their completion of coding tasks. In fact, despite
the lack of familiarity with Opiner, the participants completed the coding tasks with the 
highest accuracy  while using Opiner.}

\bf{External Validity Threats} compromise the confidence in stating whether the
study results are applicable to other groups. 
Due to the diversity of the domains where APIs can be used and developed, the
generalizability of the approach requires careful assessment. While the current
implementation of the approach applies for Java APIs, the evaluation corpus
consists of the APIs discussed in the SO threads tagged as
`Java+JSON'. We used this same dataset previously to mine and summarize reviews
about APIs~\cite{Uddin-OpinerReviewAlgo-ASE2017,Uddin-OpinionValue-TSE2019,Uddin-OpinerReviewToolDemo-ASE2017}. We observed that the dataset offers diverse perspectives about many
competing APIs. To evaluate the effectiveness of the API documentation in Opiner, we selected four coding tasks, each of which required the consultation of 
two SO posts. This decision is  based on previous research findings that developers often need to consult diverse resources to complete a coding task using an API~\cite{Ekwa-StudyUnfamiliarAPIs-ICSE2012}. 
In our previous research, we also found that developers often use a set of usage scenarios of an API in a sequence~\cite{Uddin-TemporalApiUsage-ICSE2012}. For example, we found that 
developers using the HttpClient API first established an HTTP connection, before sending/receiving messages over the HTTP connection. Such sequences of API usage scenarios 
are often not found together in SO posts. Our focus in the coding tasks was to mimic real-world development scenarios. However, it may happen that a SO post 
may still contain a complete sequence of API usage scenarios. We leave the evaluation of Opiner against such usage examples from SO as future work.     
In summary, while our
assessment of Opiner shows promising signs in this new research direction, 
the results will not carry the automatic implication that
the same results can be expected in general. Transposing the results to other
domains requires an in-depth analysis of the diverse nature of
challenges and characteristics each domain and resources can present.


%

%% file: related-work.tex
\section{Related Work}\label{sec:related-work}

Related work can broadly be divided into two categories:
\begin{inparaenum}[(1)]
\item API documentation efforts,  and 
\item Assisting development task completion using knowledge shared in developer forums.
\end{inparaenum} We summarize and compare our study with related work in \tbl\ref{tab:compareRelatedWork}. 
We discuss selected research work below.

\begin{table}[t]
  \centering
  \caption{Comparison between our study and previous related work}
   \resizebox{\columnwidth}{!}{%
    \begin{tabular}{p{1.5cm}p{4cm}|p{4cm}|p{5cm}}\toprule
    \textbf{Theme} & \textbf{Our Study} & \textbf{Prior Study} & \multicolumn{1}{l}{\textbf{Comparison}} \\ 
    \midrule
    \bf{Automatic API Documentation Efforts} & We propose two novel algorithms to automatically produce 
    API documentation from SO. Each algorithm takes as input a list of all mined 
    API usage scenarios from SO. The produced API documentation is presented using statistical metrics (e.g., star ratings of reviews) 
    and by grouping conceptually similar API usage scenarios.   & 
    Previous research mined API usage concepts and traces from the version history of clients or the source code of an API~\cite{Uddin-TemporalApiUsage-ASE2011,Uddin-TemporalApiUsage-ICSE2012,Robillard-AutomaticSuggestionProgramNavigation-FSE2005,Robillard-RecommendChangeClusters-JournalSoftwareMaintain2010a,Robillard-RepresentingConcern-PhDThesis2003}.
    Javadocs of API classes are annotated with code examples and interesting textual insights of an API from SO~\cite{Subramanian-LiveAPIDocumentation-ICSE2014,Treude-APIInsight-ICSE2016}.
     
    Techniques are proposed to add clarification to API usage using crowd-sourced knowledge~\cite{Ren-DemystifyOfficialAPIUsageDirectivesWithCorwdExample-ICSE2020}.
    & We identify API usage concepts by mining API usage scenarios from SO. We do not annotate Javadocs with our documented API usage scenarios. Instead, we present those 
    in an online API portal using innovating documentation presentation format. More than 80\% participants in our user study wished for our produced API documentation 
    to be included in official API documentation. \\
    \cmidrule{2-4}
     & We include reviews (i.e., positive/negative opinions) about code examples in our produced API documentation to give 
     information about the quality of the shared code examples.  & 
     Recent research finds that code examples shared in online forums can miss important details, 
     could be partially correct, may not be directly usable, and may contain API misuse patterns~\cite{Zhang-AreCodeExamplesInForumReliable-ICSE2018,Yang-QueryToUsableCode-MSR2016,Terragni-CSNIPPEX-ISSTA2016,Ren-DiscoverControversialDiscussions-ASE2019} 
     &  Previously, we found that developers  utilize the reviews in online forums 
     to assess the goodness of the code examples and they prefer the reviews over API official documentation~\cite{Uddin-SurveyOpinion-TSE2019}. 
     We are aware of no previous API 
     documentation effort that combine API code examples with reviews. \\
    \midrule
    \bf{Assisting Development Task Completion} & We implement and deploy our proposed algorithms and a Javadoc adaptation as Type-based documentation in an online Tool, Opiner. 
    Our two user studies show that tool is effective for developers to complete coding tasks. 
    The users preferred the API documentation produced by our two proposed algorithms over the Type-based documentation across all development scenarios. & 
    Code segments from SO and  YouTube are recommended within an IDE/search engine 
    to assist in programming tasks~\cite{Ponzanelli-PrompterRecommender-EMSE2014,Ponzanelli-CodeTube-ICSE2016}. Techniques are proposed 
    to automatically recommend and summarize quality answers to a 
    question/bug~\cite{LiSun-LearningToAnswerProgrammingQuestions-JIS2018,BowenXu-AnswerBot-ASE2017,YuanTian-APIBot-ASE2017,Campos-SearchSOPostAPIBug-CASCON2016}, and 
    to improve software query reformulation~\cite{Silva-RecommendSolutionTaskUsingSO-ICPC2019,Masud-QueryReformulationSO-EMSE2019,Masud-RACK-ICSE2017,Masud-RACK-ICSME2016}. 
    & We do not provide suggestions to complete a development task. Instead, we produce API documentation. However, in our concept-based documentation algorithm, 
    we cluster API usage scenarios that can be used to complete conceptually similar development tasks (e.g., establish HTTP connection before sending an HTTP message). Developers 
    in our user studies wrote the most correct code using least effort and least time while using Opiner, compared to while using API official documentation, SO, and/or 
    search engine. 
     \\
    \bottomrule
    \end{tabular}%
    }
  \label{tab:compareRelatedWork}%
\end{table}%

\subsection{API Documentation Efforts Using Crowd-Sourced Knowledge}
The inference of API usage patterns and properties by mining software repositores (e.g., version history of client software) has been 
an active research area for over a decade. Dynamic analysis is used on the execution traces to detect usage
patterns~\cite{Lo-MiningHierarchicalScenarioBasedSpec-ASE2009,Lo-MiningIterativePatterns-SIGKDD2007,Fahland-MiningBranchingTimeScenarios-ASE2013,Safyallah-PatternMiningDynamicAnalysis-ICPC2006}.
The techniques instrument software and collect execution traces by running the
software for relevant task scenarios. Static analysis is used on the version control
data of software to find API usage
concepts~\cite{Uddin-TemporalApiUsage-ASE2011,Uddin-TemporalApiUsage-ICSE2012,Robillard-AutomaticSuggestionProgramNavigation-FSE2005,Robillard-RecommendChangeClusters-JournalSoftwareMaintain2010a,Robillard-RepresentingConcern-PhDThesis2003}.
Unlike above research, we mine API usage scenarios from developer forum, SO.  

The automated mining of crowd-sourced knowledge from developer forums to support API documentation has broadly focused on two areas in recent years:
\begin{inparaenum}
\item Creation of API documentation and usage patterns from crowd-sourced knowledge, and 
\item Assessing the quality and feasibility of crowd-sourced developer knowledge.
\end{inparaenum}

\subsubsection{Producing API Documentation and Usage Patterns.} 
Javadocs are annotated by Baker~\cite{Subramanian-LiveAPIDocumentation-ICSE2014} to add code examples to the official documentation 
of an API type (e.g., class) from SO and by API Insight~\cite{Treude-APIInsight-ICSE2016} to add important textual insights 
about the API type. Unlike Baker and API Insight, we produce stand-alone API documentation by proposing two novel algorithms: Statistical and Concept-Based. 
API sequence (e.g., one after another or together) 
calls are mined by Gu et al.~\cite{KimGu-DeepAPILearning-FSE2016} from GitHub code repositories and by Azad et al.~\cite{Azad-GenerateAPICallrules-TOSEM2017} 
on Android code base and SO code examples. We do not produce API sequence calls in our API documentation. Instead, in our concept-based 
documentation we find API usage scenarios that are conceptually relevant to each other. Li et al.~\cite{LiXing-LeveragingOfficialContentSoftwareDocumentation-TSC2018} 
propose CnCxL2R, a software documentation recommendation 
engine by incorporating the content of the official documentation with social context from SO. 
Souza et al.~\cite{Souza-CookbookAPI-BSSE2014,Souza-BootstrapAPICodeBookSO-IST2019} propose a semi-automatic approach based 
on topic modeling to build cookbooks for APIs by organizing 
knowledge available in SO. Topic modeling was also applied by Campbell et al.~\cite{Campbell-DeficientDocumentationDetection-MSR2013} 
on a combined dataset of SO, and PHP and Python official documentation to determine missing topic coverage in the official documentation. 
Ren et al.~\cite{Ren-DemystifyOfficialAPIUsageDirectivesWithCorwdExample-ICSE2020} 
develop a text mining approach to discover API misuse scenarios in SO. 
They extract erroneous code examples, patches, related and confusing APIs from the misuse scenarios and construct demystification reports. 
The reports can be 
used to help developers understand the official API usage directives. 
Unlike the above work, we also 
consider reviews about code examples while producing the documentation. In our previous surveys of 178 software 
developers, we found that developers consider the combination of code examples and reviews about the examples as a form of API documentation and they 
value such documentation over official documentation. 

To produce API documentation from crowd-sourced forum, we first need to link code examples to APIs.
A number of techniques automatically infer code terms (e.g., API method, class)
in textual contents and code examples of forum posts (see \cite{Dagenais-RecoDocPaper-ICSE2012a},
\cite{Subramanian-LiveAPIDocumentation-ICSE2014}, \cite{Rigby-CodeElementInformalDocument-ICSE2013}, \cite{YeDeheng-ExtractAPIMentions-ICSME2016},
\cite{Phan-StatisticalLearningFQNForums-ICSE2018},
and \cite{Ma-APINERSODeepLearning-TSE2019}).
Unlike above techniques, we associate a code example to an
API mentioned in the textual contents of the forum post, about which the code example is provided. Therefore, we do not link
\it{all} the different APIs whose classes are used in the code example or all different API/code elements that are mentioned in the textual contents.
In our previously published paper~\cite{Uddin-MiningAPIUsageScenarios-IST2020}, we showed that this approach is better suited to produce
task-centric~\cite{Carroll-MinimalManual-JournalHCI1987a,DeSouza-DocumentationEssentialForSoftwareMaintenance-SIGDOC2005}
API documentation from SO, where an API is of particular focus to complete a task. 

\subsubsection{Assessing Quality and Documentation Feasibility of Forum Contents} 
The recent focus on API documentation creation from SO and online contents is motivated by previous research findings on analyzing the 
feasibility of SO to support API documentation and program comprehension. Parnin and Treude~\cite{Parnin-MeasuringAPIDocumentationWeb-Web2SE2011} 
analyzed to what extent the methods of an API (jQuery) are documented on the Web. They analyzed 
1730 search results and find that 87.9\% of the methods are mainly covered in online blogs and tutorials. Jiau and Yang~\cite{Jiau-FacingInequalityCrowdSourcedDocumentation-SENOTE2012}
find that API classes with with lower documentation can be covered by another class with more documentation in SO, where both are found under same inheritance.  
Delfim et al.~\cite{Delfim-RedocummentingAPIsCrowdKnowledge-JournalBrazilian2016} find that developers in SO provide more content for debugging tasks then for how-to-do tasks.
Sunshine et al.~\cite{Sunshine-APIProtocolUsability-ICPC2015}  find that 
developers spent significant time in online forums to seek information about API protocol usage. 
Studies found code examples from different 
programming languages can have different usability factors in SO~\cite{Yang-QueryToUsableCode-MSR2016}, 
some Android API classes in the shared code can be more challenging to use than other~\cite{Wang-APIsUsageObstacles-MSR2013}, 
but there is sub-linear relationship between the Android class popularity in Android apps 
and the requests for their documentation in SO~\cite{Kavaler-APIsUsedinAndroidMarket-SOCINFO2013}. The encouraging documentation coverage in the Web, 
in participate in SO, have motivated us to produce our API documentation from SO. 
Zhang et al.~\cite{Zhang-ReadingAnswerSONotEnough-TSE2019} find that majority of comments posted in SO are informative and thus they can enhance the quality of their associated 
answers. They propose to include comments into knowledge seeking and documentation resources, which traditionally have been only using accepted answers from SO.
Following suggestions from Zhang et al.~\cite{Zhang-ReadingAnswerSONotEnough-TSE2019},  
we analyze both answers and comments to code examples while producing our API documentation.

The encouraging findings on the feasibility and availability of crowd-sourced knowledge have motivated researchers to also 
analyze the quality of the shared knowledge. A number of recent research papers~\cite{Zhang-AreCodeExamplesInForumReliable-ICSE2018,Yang-QueryToUsableCode-MSR2016,Terragni-CSNIPPEX-ISSTA2016} 
warn against directly copying code from SO, because such code can have potential bugs or 
misuse patterns~\cite{Zhang-AreCodeExamplesInForumReliable-ICSE2018} and that such code 
may not be directly usable (e.g., not compilable)~\cite{Yang-QueryToUsableCode-MSR2016,
Terragni-CSNIPPEX-ISSTA2016}. Mondal et al.~\cite{Mondal-SOIssueReproducability-MSR2019} analyze the reproducibility of code example shared in SO questions. They find that questions with 
reproducible problem in the code example are three times more likely to be answered. 
Mastrangelo et al.~\cite{Mastrangelo-JavaUnsafeAPIs-OOPSLA2015} investigate the spread of unsafe Java code that are used as ``backdoors'' in Java runtime to facilitate 
high-performance system level code in Java. They analyze 74GB of compiled Java code, spread over 86,479 Java archives. They find that 25\% of the Java archives use such unsafe code.
Zhang et al.~\cite{Zhang-ObsoleteAnswerSO-TSE2019} analyze answers with obsolete solutions in SO. They find that more than 50\% of those obsolete answers were obsolete even during 
the time of their creation. They find that answers to questions related to node.js, ajax, android, and objective-c are more likely to become obsolete. 
Tools and techniques are also proposed to detect high quality posts and answers~\cite{Ponzanelli-ImproveLowQualityPostDetect-ICSME2014}.
Recently, Ren et al.~\cite{Ren-DiscoverControversialDiscussions-ASE2019} propose a technique to discover answers that are being criticized in SO for having erroneous 
solutions. They identify such critique posts as controversial discussions. They then summarize the controversial discussions per problematic answer. 
In our current Opiner API documentation framework, we inform developers of problems in the code examples by showing them negative opinions provided 
against the code examples in the comments to the answers. The developers in our user studies mentioned that such insights are very useful. The produced 
API documentation can be made more useful by highlighting controversial discussions about a code example, by utilizing the techniques from Ren et al.~\cite{Ren-DiscoverControversialDiscussions-ASE2019}.

Ragkhitwetsagul et al.~\cite{Ragkhitwetsagul-ToxicCodeSO-TSE2018} identify toxic code in SO as the code examples that are outdated or that violate original software 
license. They conduct a survey of 201 high-reputation SO users and find that they rarely check for license violation. 
Violation of licenses is a common problem for code examples shared in other platforms, such as Android Ecosystem (see Mlouki et al.~\cite{Mlouki-LicenseViolationAndroid-SANER2016}). 
All documented API usage scenarios in Opiner can be further enhanced by the information of license that need to be adopted before the reuse of a code example. This can 
help reduce the above license violation problems.
 
%
%
%
%
%
%
%
%
%

\subsection{Assisting Development Task Completion Using Crowd-Sourced Knowledge}
A number of tools are developed to offer context-aware code completion information into the IDE of a developer based on crowd-sourced knowledge. 
Zilberstein and Yahav~\cite{Zilberstein-Codota-ONWARD2016} proposed an approach 
to leverage code examples from forum posts to support auto code completion features in IDEs.
They built Codota, a tool that given a code context from the IDE, 
automatically searches forum posts to find most relevant code examples that can be relevant to the code context.
Ponzanelli et al.~\cite{Ponzanelli-PrompterRecommender-EMSE2014} developed an
Eclipse Plug-in that takes into account the source code in a given file as a
context and use that to search SO posts to find relevant discussions
(i.e., \it{code to relevant information}). Campbell and Treude~\cite{Campbell-NLP2Code-ICSME2017} develop NLP2Code that integrates with a developer's IDE to provide him with 
a content assist feature to close the vocabulary gap between the needs of the developer and the code snippet meta data. 
While the above tools extend IDEs, Opiner is a stand-alone portal to show 
API documentation from SO. Unlike the above tools that only use code example/discussions, Opiner use code example and reviews.

Developers query for technical contents online. A number of techniques are proposed to provide better results to an input technical query by 
leveraging crowd-sourced knowledge.  Campos et al.~\cite{Campos-SearchSOPostAPIBug-CASCON2016} propose an approach to find fixes to API-related bugs bugs by matching code snippet that are being 
debugged against related snippets in SO. To find the API-related bugs, they used OHLOH code search engine containing potentially buggy API method calls. 
Rahman et al.~\cite{Masud-QueryReformulationSO-EMSE2019,Masud-RACK-ICSE2017,Masud-RACK-ICSME2016,Masud-NLP2API-ICSME2018} 
propose RACK and NLP2API that suggest a list of relevant API classes for a natural 
language query intended for code search. The technique exploits keyword-API associations from SO posts to reformulate the queries. 
Silva et al.~\cite{Silva-RecommendSolutionTaskUsingSO-ICPC2019} propose a tool CROKAGE (Crowd Knowledge Answer Generator) that 
takes as input a natural language query of the description of a programming task and outputs solutions to the task by offering code examples 
and explanations. The query is expanded with relevant API classes from SO. Li et al.~\cite{LiSun-LearningToAnswerProgrammingQuestions-JIS2018} develop QDLinker, a deep-learning-to-answer framework to answer programming questions with software 
documentation. The tool learns from the discussions in SO to bridge the semantic gap between the types of questions asked and software documentation. 
Based on the learning, the tool can effectively answer a programmer question with links to software documentation. Unlike the above techniques, 
all the documentation of an API in Opiner can be searched by simply searching by the API name. 
Thus Opiner can be extended with above techniques, so that developers can search Opiner API documentation using 
natural language query. In fact, during our surveys, this is also one of the features that developers wished Opiner could have in future extensions.

Recently, bots are developed to assist developers in diverse development tasks by utilizing crowd-sourced knowledge. Xu et al.~\cite{BowenXu-AnswerBot-ASE2017} develop AnswerBot that takes as input a natural language description of a programming task. The tool 
finds questions relevant to the query and summarizes the useful answers from those questions. Tian et al.~\cite{YuanTian-APIBot-ASE2017} develop APIBot to answer API questions given API documentation as an input. The Bot is built by adapting SiriusQA, a cross-domain 
intelligent personal assistant. The engine achieve a Hit@5 score of 0.706 to answer the questions. Future extensions of the bots can also target the curated API documentation in Opiner.

Finally, contents in SO can be better used, if the underlying contexts can be better processed. 
Ahsanuzzaman et al.~\cite{Ahsanuzzaman-ClassifySOPost-SANER2018,Ahasanuzzaman-ClassifyIssueSO-EMSE2019} develop machine learning classifier to automatically label SO questions 
describing an issue or not. Terragni et al.~\cite{Terragni-CSNIPPEX-ISSTA2016} investigate the compilability issues with the code snippets shared in SO. They developed CSnippEx 
that can automatically convert a shared Java code into a compilable code by resolving external dependencies, import statements and syntax errors. Both of these 
techniques can help extend the API documentation in Opiner, e.g., by producing compilable code using CSnippEx.

%% file: summary.tex
\section{Conclusions}\label{sec:summary}
The rapid growth of online developer forum posts 
encourages developers to harness the knowledge of the crowd in their daily development activities. 
The enormous volume of such insights distilled into the API code examples presents challenges to the developers to gain quick and actionable insights into the usage of an API.
In this paper, we propose two novel algorithms to automatically produce API documentation from SO: Statistical and Concept-based. Each algorithm takes as input a list of mined API usage 
scenarios from SO. Statistical algorithm produces visualized metrics, while concept-based algorithm clusters conceptually relevant code examples. 
An API usage scenario consists of a code example, a textual task description, and the reviews towards the code example. A review is an opinionated sentence with positive 
or negative sentiments. In
three user studies involving Opiner, we observed that:
\begin{enumerate}
\item The two proposed algorithms, Statistical and Concept-based Documentation are preferred over a traditional Javadoc styled Type-based documentation by the 
study participants across all the four development scenarios.
\item Developers spent the least time and effort while coding solutions to development tasks 
using Opiner compared to formal and informal documentation. Moreover, the accuracy of their solutions is highest using Opiner. 
\item  More than 80\% of developers considered that Opiner offer improvements over both formal
and informal API documentation, and wished to use Opiner in their daily or weekly development
activities.
\end{enumerate} The findings from our study can guide \begin{inparaenum}
\item \it{API Authors and Documentation Writers} to improve the official documentation of their APIs, 
\item \it{Software Developers} by offering them a new and potentially better API documentation format than traditional Javadocs, and 
\item \it{Researchers} to study new ways of documenting API usage scenarios from developer forums.
\end{inparaenum} 
 
Our future work focuses on the better understanding of the interplay between comments and answers. 
In particular, we analyze how we can better support API documentation by automatically incorporating insights from comments. For example, comments 
can be useful to know whether a code example is applicable to a given API version or whether the code has any other compatibility or buggy issues. 
Currently, we simply show all comments towards a code example by grouping the sentences from the comments into three polarity classes: 
sentences containing positive, negative or neutral sentiments. In addition, we can determine whether suggestions from comments are already incorporated into the 
answers. This can be checked by comparing the different suggested and approved edits to an answer and whether such edits were made after a constructive comment 
was provided. With such intelligent analyses, we can then reduce the number of comments that are already addressed.  In addition, we will focus on the 
following areas: \begin{inparaenum}[(1)]
\item Automatic mining and documentation of usage scenarios from multiple
developer forums,
\item Addition of improved search capabilities to facilitate the discovery of
usage summaries in Opiner, and
\item Incorporation of a continuous feedback learner in the Opiner documentation engine that
can automatically learn from users feedback to improve the documentation.
\end{inparaenum} 